\documentstyle[preprint,prc,aps]{revtex}
\def\pmb#1{\setbox0=\hbox{#1}%
  \kern-.02em\copy0\kern-\wd0
  \kern.04em\copy0\kern-\wd0
  \kern-.02em\raise.04em\box0 }
\def\bolddot{{\pmb{$\cdot$}}}
\begin{document}
\tighten
\preprint{IU/NTC 96--03}
\title{Relativistic Mean-Field Theory and the High-Density
Nuclear Equation of State}
\author{Horst M\"uller and Brian D. Serot}
\address{Physics Department and Nuclear Theory Center\\
        Indiana University,\ \ Bloomington,~Indiana\ \ 47405}
\vskip1in
\date{\today}
\maketitle

\begin{abstract}
The properties of high-density nuclear and neutron matter are
studied using a relativistic mean-field approximation to the 
nuclear matter energy functional.
Based on ideas of effective field theory, nonlinear interactions between the
fields are introduced to parametrize the density dependence of the energy
functional. 
Various types of nonlinearities involving scalar-isoscalar ($\sigma$),
vector-isoscalar ($\omega$), and vector-isovector ($\rho$) fields are
studied.
After calibrating the model parameters 
at equilibrium nuclear matter density, the
model and parameter dependence of the resulting equation of state is
examined in the neutron-rich and high-density regime. 
It is possible to build different models that reproduce the same observed 
properties at normal nuclear densities, but
which yield maximum neutron star masses that differ by more than
one solar mass.
Implications for the existence of kaon condensates or quark cores in
neutron stars are discussed.
\end{abstract}

\vspace{20pt}
\pacs{PACS numbers: 21.65.+f, 26.60.+c, 97.60.Jd}

%
\section{Introduction}
Numerous calculations have established that
relativistic mean-field models provide a realistic description of the 
bulk properties of finite nuclei and nuclear matter \cite{SW86,BDS92}.
In addition to this successful low-energy phenomenology, these models
are often extrapolated into
regimes of high density and temperature to extract the nuclear equation of
state (EOS), which is the basic ingredient in many astrophysical 
applications and in microscopic models of energetic nucleus--nucleus 
collisions.

Based on the original version of Walecka \cite{WALECKA74} and its
extensions \cite{BOGUTA77,SEROT79}, relativistic mean-field
models generally involve the interaction of Dirac nucleons with neutral
scalar and vector mesons and with isovector $\rho$ mesons.
One of the key observations in their success is that to provide sufficient
flexibility, nonlinear self-interactions for the scalar meson must be
included 
\cite{BDS92,BOGUTA77,REINHARD86,FPW87,FURNSTAHL89,BODMER89,GAMBHIR90,%
FURNSTAHL93}.
Since these models were proposed to be renormalizable, the scalar 
self-interactions are limited to a quartic polynomial, and scalar--vector
and vector--vector interactions are not allowed \cite{BOULWARE70}.
One of the motivations for renormalizability, as discussed in 
Walecka's seminal paper, is that once the model parameters are
calibrated to observed nuclear properties, one can extrapolate into
regimes of high density or temperature without the appearance of new,
unknown parameters.

An alternative approach is inspired by {\em effective} field theories, 
such as chiral perturbation theory \cite{WEINBERG68,DASHEN69}, which 
successfully describes the low-energy phenomenology of hadronic 
Goldstone bosons \cite{GL8485,ChPT94}.
Although a lagrangian usually serves as the starting point,
the meson and baryon fields are no longer considered elementary, 
and the constraint of renormalizability is dropped.
This has several important consequences.
First, there is no reason to restrict meson self-interactions to a simple
quartic polynomial in the scalar field; on the contrary, one should 
include all interaction terms that are consistent with the underlying
symmetries of QCD.
Second, since there are an infinite number of coupling constants, one must
find suitable expansion parameters for the systems under consideration,
and one must develop a systematic truncation scheme for the effective
theory to have any predictive power.
Third, extrapolation of calculated results into new regimes of the
physical parameters becomes problematic, because the truncation scheme may
break down, and predictions can become sensitive to unknown parameters.

Within the framework of effective field theory, mean-field models of nuclear
structure and the EOS must be interpreted in a new context.
One important observation is that near normal nuclear density, the mean 
scalar and vector fields (or nucleon self-energies), 
which we denote as $\Phi$ and $W$, are large on
nuclear energy scales but are small compared to the nucleon mass $M$
and vary slowly in finite nuclei.
This implies that the ratios $\Phi /M$ and $W/M$ and the gradients
$|\nabla \Phi |/M^2$ and $|\nabla W |/M^2$ are useful expansion parameters.
The assumption of ``naturalness'' in effective field theory is 
also important.
Naturalness implies that the coefficients of the various terms in the
lagrangian, when expressed in appropriate dimensionless form,
should all be of order unity.
When combined with meaningful expansion parameters,
this means that one can anticipate the approximate magnitude of mean-field 
contributions to the energy (at least up to moderate nuclear
densities) and thereby motivate a suitable truncation scheme; if the 
coefficients are natural, the omitted terms will be numerically 
unimportant.\footnote{It has also been shown recently that the naturalness
assumption is consistent with dimensional counting in chiral perturbation
theory \protect\cite{FRIAR96,VMD96}.}
Naturalness also implies that one should
include {\em all\/} possible terms (that is, those allowed by the 
symmetries) through a given order of truncation; it is {\it unnatural\/} 
for some coefficients to vanish without a relevant symmetry argument.

From this point of view, it is difficult to justify nuclear mean-field 
models that include only scalar self-interactions
\cite{BOGUTA77,REINHARD86,FPW87,GAMBHIR90}, and
recently, generalizations that also include quartic self-interactions for 
the neutral vector meson have been 
discussed \cite{BODMER89,BODMER91,GMUCA92}.
Moreover, a new analysis involving all meson self-interactions through 
fourth order in the isoscalar
scalar and vector fields has been performed \cite{FST96}.
These extensions give rise to additional model parameters (coupling 
constants) that must be constrained by calibrating to observed nuclear
properties.
For the truncation at fourth order to be sensible, the parameters so obtained
should exhibit naturalness.

Although it is possible to discuss effective hadronic field theory from
the point of view of a lagrangian, as above, the expansion in powers of the
mean fields is a low-density expansion, and it is hard to justify the 
neglect of many-body corrections, which are known to be relevant in nuclear
structure and in the EOS.
Alternatively, one can consider this expansion at the level of an energy
functional or effective action \cite{FST96}.
In such a formulation of the relativistic nuclear 
many-body problem, the central object is an energy functional of scalar and 
vector densities (or more generally, vector four-currents)
\cite{DREIZLER90,SPEICHER92,SCHMID95}.
Extremization of the functional gives rise to Dirac equations 
for occupied orbitals
with {\it local\/} scalar and vector potentials, 
not only in the Hartree approximation, but in the general case as well.
Rather than work solely with the densities,
one can introduce auxiliary variables corresponding to the local potentials,
so that the functional depends also on mean meson fields.
The resulting field equations have the same form as in a Dirac--Hartree 
calculation \cite{FST96}, but correlation effects can be included, 
{\it if\/} the proper energy functional can be found.
This procedure is analogous to the well-known Kohn--Sham \cite{KOHN65}
approach in density-functional theory, with the local meson fields playing 
the role of (relativistic) Kohn--Sham
potentials; by introducing nonlinear couplings between these fields, one can
implicitly include additional density dependence in the potentials.
Thus the nonlinear meson interaction
terms simulate more complicated physics, such as one- and two-pion exchange,
or vacuum-loop corrections, which might be calculated directly in a more
microscopic many-body approach \cite{GMUCA92,SCHMID95}.
The fields (and their gradients) again serve as useful expansion parameters
at moderate density, so the nonlinear interaction terms can be truncated,
leaving a finite number of unknown couplings.

Rather than focus on the calculation of the nonlinear couplings from an
underlying effective lagrangian, we wish to concentrate instead on how well
the energy functional can be calibrated by fitting the couplings to observed
nuclear properties, and on the limitations on the extrapolation of the
resulting EOS into the high-density regime.
In general, even with a significant truncation, the number of unknown
couplings exceeds the number of normalization conditions, which we take
to be five properties of infinite nuclear matter: the equilibrium
density and binding energy ($\rho_0$, $-e_0$), the nucleon effective
(or Dirac) mass at equilibrium ($M^\ast_0$), the compression modulus
($K_0$), and the bulk symmetry energy ($a_4$).
(Experience has shown 
that an accurate reproduction of these five properties leads to realistic
results when the calculations are extended to finite 
nuclei \cite{REINHARD86,FPW87,GAMBHIR90,FURNSTAHL93,FST96}.)
Thus families of models can be generated
which describe exactly the same nuclear matter properties at equilibrium
\cite{BODMER91}.
We then investigate the differences in the high-density EOS
predicted by models within a given family.
This question is important for astrophysical applications, particularly in
the study of neutron stars; it will be difficult to deduce the existence of
``exotic'' neutron-star structure (for example, hyperonic matter, kaon 
condensates, or quark cores) unless the EOS of the more mundane components 
(neutrons, protons, and electrons) is well constrained.

If our truncation of the energy functional is motivated by low-density
behavior, why should we have any confidence at all in a high-density 
extrapolation?
This is indeed the crucial question, and we are not attempting here to 
justify such an extrapolation; we are merely recognizing that this procedure 
is often used in neutron-star calculations, even recent ones, 
without any mention of the implicit assumptions about the absence of 
additional contributions at high density
\cite{FP81,GLENN92,MARU94,KNORR95,PRAK95a,PRAK95b,FUJII96}.
We therefore feel it is timely to investigate quantitatively the 
uncertainties in the extrapolated equation of state.

We begin the theoretical analysis with a model that contains meson--meson
and meson self-interactions described by an {\em arbitrary\/} finite
polynomial in the fields.
We find that both the asymptotic (high-density) limit of the EOS and the
approach to this limit ({\em i.e.}, the ``stiffness'') are model dependent.
In particular, one can construct models with the {\em same\/} equilibrium 
nuclear matter properties that yield high-density equations of state 
ranging all the way from the causal limit ($p={\cal E}$) to 
one that resembles a free relativistic gas ($p = {\cal E}/3$) 
\cite{BODMER89,BODMER91}.
(Here $p$ is the pressure and $\cal E$ is the energy density.)

As an explicit example, we consider a model that includes self-interactions 
for the isoscalar scalar and vector mesons and for the $\rho$ meson up
to fourth order in the fields. 
To our knowledge, this is the first time nonlinear terms in the $\rho$ meson 
mean field have been included.
(Note that the $\rho$ field enters here as an {\em effective\/} field whose
purpose is to parametrize the isospin dependence; thus, fundamental questions
about causal propagation \cite{VELO69,VELO73} 
and spin mixing \cite{OP63} are not relevant.)
To provide a quantitative measure of the variations in the EOS, we compute
neutron star masses, which turn out to be sensitive to the model types
and to changes in the parametrizations, {\em even for models that reproduce
the same equilibrium nuclear matter properties}.
In some cases, variations in the calculated maximum mass are more than one
solar mass.

These results lead us to two basic conclusions.
First, existing methods for calibration of the EOS at normal density are
{\em not\/} sufficient to provide a satisfactory extrapolation into the 
density regime relevant for neutron stars.
The basic problem rests with the quartic neutral vector meson 
($\omega$) interaction,
which produces major modifications in the high-density behavior.
We discuss the existing situation regarding the specification of this term
and prospects for improved calibration.
Second, we find that if the quartic $\omega$ term can be accurately 
determined, additional higher-order terms (and the quartic $\rho$ term)
produce relatively minor changes in the neutron star mass.
This occurs because the $W^4$ term softens the high-density EOS so
completely that additional interactions have little effect, at least in
the density regime relevant for neutron stars.

These conclusions have important implications, because the contributions 
to the EOS from the neutrons and protons are commonly believed to be the best
understood part of the physical input necessary to describe a dense 
stellar object.
Without well-constrained results from this part of the EOS, it will
be impossible to determine the importance of additional dynamics,
for example, the transition from nuclear matter to quark matter in
the interiors of neutron stars \cite{COLLINS75,BAYM76,GLENN92}, or the 
role of strangeness in the form of hyperons \cite{GLENN85,ELLIS91} or 
a kaon condensate \cite{POLITZER91}.
As an example of how uncertainties in the basic nuclear EOS can influence 
these interesting effects, we study the transition from hadronic matter to 
quark matter using a simple model \cite{BAYM76,SW86} and show the possible 
variations in the results. 

The outline of this paper is as follows:
In Sec.~II, we present the general model, which involves an arbitrary
number of nonlinear meson interactions, and derive the EOS.
Based on this general model, Section III is devoted to the high-density 
limit of the EOS.
In Sec.~IV, we apply our model to neutron stars. 
For the quantitative analysis, we initially include self-interactions 
up to fourth order in the $\rho$ and $\omega$ fields and then investigate
the consequences of sixth-order and eighth-order $\omega$ self-interactions. 
We also briefly discuss the parameter and model dependence
of the transition to quark matter in the central region of the star.
Section V contains a short summary and our conclusions.
%

%
\section{The Nuclear Equation of State}
We describe the nuclear equation of state using a relativistic approach
involving valence Dirac nucleons and effective mesonic degrees of freedom, 
which are taken to be neutral scalar and vector fields, plus the isovector
$\rho$ meson field.
Rather than focus directly on a lagrangian, we consider instead as a 
starting point an {\em effective action}
\begin{eqnarray}\Gamma=\Gamma [\phi,V_{\mu},{\bf b}_{\mu}] \ ,
			   \label{eq:Action1}
                      \end{eqnarray}
which is a functional of the meson fields denoted by
$\phi$, $V_\mu$, and ${\bf b}_\mu$ for the scalar, vector-isoscalar,
and vector-isovector field, respectively.
In principle, this functional can be calculated in a many-body
approach based on a Lagrangian for the nucleon--nucleon interaction, 
or its general form might be obtained from an underlying theory.
Here we will be satisfied to parametrize the effective 
action, calibrate it as accurately as we can to observed nuclear properties,
and then examine the predicted high-density equation of state.

The effective action is related to the thermodynamic 
potential $\Omega$ \cite{FETTER71} by
\begin{eqnarray} i\beta\Omega=\Gamma [\phi,V_{\mu},{\bf b}_{\mu}]\ ,
			   \label{eq:omega1}
                      \end{eqnarray}
where the fields are determined by the general thermodynamic principle that
they should make $\Omega$ stationary:
\begin{eqnarray}
{\partial \Gamma\over \partial \phi}\,=\,
{\partial \Gamma\over \partial V_{\mu}}\,=\,
{\partial \Gamma\over \partial {\bf b}_{\mu}}\,=\, 0 \ .
			   \label{eq:extremum}
                      \end{eqnarray}
A basic property of the functional is that it reflects the underlying
symmetries \cite{ITZYKSON80}. 
Thus, if we assume that the system possesses two conserved charges, namely,
baryon number $B$ and the third component of total isospin $I_3$, this
gives rise to two chemical potentials $\mu$ and $\nu$:
\begin{eqnarray}\Omega \equiv\Omega(\beta,\mu,\nu) \ ,
			   \label{eq:omega2}
                      \end{eqnarray}
with
\begin{eqnarray}
    B&=&\int d^3x \,\rho=-{\partial\Omega \over \partial \mu}\ ,
        \label{eq:bdef}\\
    I_3&=&{1\over 2}\int d^3x \,\rho_3=-{\partial\Omega \over \partial \nu} 
          \ ,\label{eq:idef}
                      \end{eqnarray}
$\beta$ the inverse temperature, and $\rho_3 \equiv \rho_{\rm p}
-\rho_{\rm n}$.
Note that the fields can be held fixed in evaluating the partial
derivatives in Eqs.~(\ref{eq:bdef}) and (\ref{eq:idef}) by virtue of
the extremization conditions (\ref{eq:extremum}).

In the field theoretical context \cite{ILIOPOULOS75}, one can show that
the effective action can be expanded as a power series in gradients of
the fields.
Thus, after taking the zero-temperature limit, one usually writes
\begin{eqnarray}\Gamma[\phi,V_{\mu},{\bf b}_{\mu}]
                =\int d^4 x \Bigl[-{\cal V}_{\rm eff}(\phi,V_{\mu},
        {\bf b}_{\mu})
        +{\cal Z}(\phi,V_{\mu},{\bf b}_{\mu},
\partial_{\nu}\phi,\partial_{\nu} V_{\mu},\partial_{\nu}
        {\bf b}_{\mu})\Bigr]
             \ , \label{eq:action2}
                      \end{eqnarray}
where the second term vanishes in a uniform
system, and the dependence on the chemical potentials has been suppressed.
For the effective potential, we make the following {\em Ansatz\/}:
\begin{eqnarray}{\cal V}_{\rm eff}(\phi,V_{\mu},{\bf b}_{\mu}; \mu, \nu )
              &=&
       {1\over 2}\,m_{\rm s}^2 \phi^2
      -{1\over 2}\,m_{\rm v}^2 V_{\mu} V^{\mu}
          -{1 \over 2}\, m_{\rho}^2{\bf b_{\mu}}
                             \bolddot{\bf b}^{\mu}\nonumber\\
      & &\null +i\,{\rm tr} \ln Z_{\psi}(\mu, \nu)
               -i\,{\rm tr} \ln Z_{\psi}(0, 0)\nonumber\\
   & &\null + \Delta{\cal V}(\phi,V_{\mu}V^{\mu},
          {\bf b_{\nu}}\bolddot{\bf b}^{\nu}; \mu , \nu)
			\ ,   \label{eq:Pot1}
                      \end{eqnarray}
with a nonlinear potential
\begin{eqnarray}
       \Delta{\cal V}(\phi,V_{\mu}V^{\mu},
            {\bf b_{\nu}}\bolddot{\bf b}^{\nu}; \mu , \nu)
       \equiv -\sum_{i,j,k}a_{ijk}(\mu , \nu)\,\phi^i\, (V_{\mu}V^{\mu})^j\,
			   ({\bf b_{\nu}}\bolddot{\bf b}^{\nu})^k
			   \label{eq:Pot2}
                      \end{eqnarray}
that contains at least three powers of the fields: $i+2j+2k \geq 3$.
The fermionic contributions are represented by
a one-body term ${\rm tr}\ln Z_{\psi}$ (with the appropriate zero-density
subtraction \cite{FTS95})
and by terms where the fermions have been 
``integrated out'', which results in a (generally nonanalytic) dependence
of the mesonic coefficients $a_{ijk}$ on the chemical potentials
$\mu$ and $\nu$.
There will also be contributions to the $a_{ijk}$ that are independent
of the chemical potentials; these arise from integrating out heavy degrees
of freedom and vacuum loops.
Thus ${\cal V}_{\rm eff}$ contains explicit contributions only from valence 
nucleons and classical meson fields.

The fermionic part ${\rm tr}\ln Z_{\psi}$ is obtained by evaluating the
trace of the kernel
\begin{eqnarray}
 K(\mu , \nu ) = 
(i\partial^\mu -g_{\rm v} V^\mu
 -{1\over 2}\,g_{\rho}{\hbox{\boldmath$\tau$}}\bolddot{\bf b}^\mu )\gamma_\mu
      + \mu \gamma^0 + {1\over 2}\nu\tau_3 \gamma^0 
                    - (M - g_{\rm s} \phi) 
                      \label{eq:Dirac}\end{eqnarray}
using Dirac wave functions calculated in the presence of static
background fields.
The subtraction removes contributions from negative-energy states, which
are already included implicitly in the nonlinear parameters $a_{ijk}$.
(See Ref.~\cite{FTS95} for an analogous calculation.)
The values of the fields are determined by extremization at the
given values of $\mu$ and $\nu$.

The potential of Eq.~(\ref{eq:Pot1}) represents an effective field
theory for the interacting nucleons.
Although the exact form of the effective potential is not known, we 
introduce the meson mean fields as relativistic Kohn--Sham 
potentials \cite{KOHN65} and
consider the valence nucleons moving in the resulting local fields.
The nonlinear interactions of the fields generate implicit density dependence
above and beyond that arising from the couplings in Eq.~(\ref{eq:Dirac}).
Thus the series in Eq.~(\ref{eq:Pot2}) can
be interpreted as a Taylor series parametrization of the unknown part of the 
effective potential, which includes the effects of nucleon 
exchange and correlations,
as well as contributions from other mesons and the quantum vacuum.
(See the discussion in Refs.~\cite{FST96,FTS95}.)

Although the couplings $a_{ijk}$ generally depend on the
chemical potentials,
experience with calculations for finite nuclei and nuclear matter, together
with explicit computations of exchange and correlation corrections
\cite{RBBG}, implies that mean fields and {\em constant\/} couplings
$a_{ijk}$ provide an adequate
(albeit approximate) parametrization of these many-body effects.
Thus we consider the $a_{ijk}$ as constants in the sequel and leave
the study of their dependence on $\mu$ and $\nu$ as a topic for future
investigation.
Moreover, at low densities and temperatures, the mean meson fields are 
small compared to the nucleon mass, and so provide useful expansion 
parameters \cite{FST96}.
Thus, in practice, the series in Eq.~(\ref{eq:Pot2}) can
be truncated at some reasonable order, and the relevant question in this
paper is how far one can extrapolate the truncated potential into the
high-density regime.

In principle, the unknown coefficients (coupling constants) can be 
constrained by imposing chiral symmetry and other symmetries of the
underlying QCD, such as broken scale invariance.
(Lorentz covariance and isospin symmetry are
already incorporated explicitly.%
\footnote{Note that since
the energy functional is an effective functional, 
we presently know of no reason to exclude terms that explicitly
contain the medium four-velocity $u^\mu$, 
such as $u^\mu V_\mu V^\nu V_\nu$.
This issue will be considered in a later publication.\/}%
)
As has been discussed recently, however \cite{FURNSTAHL93,FST96}, if one
assumes a {\em nonlinear\/} realization of the chiral symmetry for the
pions and nucleons \cite{GEORGI84}, the meson interaction terms are 
essentially unconstrained.\footnote{%
Broken scale invariance leads to restrictions on the purely scalar
interactions, as shown in Refs.~\protect\cite{HEIDE94} and
\protect\cite{FTS95}, but we will
not consider these limitations here.
As discussed in the next section, the details of the scalar dynamics
do not play a major role in our analysis.}
We therefore take the couplings as free model parameters in our approach.
Obviously, an infinite number of normalization conditions is generally
needed to fix their values. 
Since this is not feasible in practice, we
terminate the summation at the finite values $i_{\rm max}$,
$j_{\rm max}$, and $k_{\rm max}$.

According to Eq.~(\ref{eq:extremum}), the thermodynamic potential $\Omega$ 
must be stationary with respect to changes in the fields for
fixed values of the proton and neutron chemical potentials
\begin{eqnarray}
\mu_{\rm p}&\equiv&\mu+{\nu\over 2}=
		 (k_{\rm Fp}^2+M^*{}^2)^{1/2}+W+{1\over 2}R \ ,
      \label{eq:mupdef}\\
\mu_{\rm n}&\equiv&\mu-{\nu\over 2}=(k_{\rm Fn}^2
                 +M^*{}^2)^{1/2}+W-{1\over 2}R 
      \label{eq:mundef}\ .
\end{eqnarray}
Here, following Bodmer \cite{BODMER91}, we define the scaled meson fields
$\Phi \equiv g_{\rm s} \phi$, $W \equiv g_{\rm v} V_0$, and
$R \equiv g_\rho b_0$, with $b_0$ the timelike, neutral part of the $\rho$
meson field;
the effective nucleon mass is $M^* \equiv M-\Phi$.
(We work in the rest frame of the infinite matter, where the spatial parts
of the vector fields vanish.)
The Fermi momenta for protons ($k_{\rm Fp}$) and neutrons $(k_{\rm Fn})$
are related to the conserved baryon density
\begin{eqnarray}
\rho={1\over 3\pi^2}\Bigl(k_{\rm Fp}^3+k_{\rm Fn}^3\Bigr)
\end{eqnarray}
and isovector density
\begin{eqnarray}
\rho_3={1\over 3\pi^2}\Bigl(k_{\rm Fp}^3-k_{\rm Fn}^3\Bigr)\, .
\end{eqnarray}

Using Eqs.~(\ref{eq:omega1}), (\ref{eq:bdef}), and (\ref{eq:idef}),
together with relations (\ref{eq:mupdef}) and (\ref{eq:mundef})
for the chemical potentials,
it is  straightforward to eliminate
the chemical potentials in favor of the densities
and to compute the pressure $p$ and the energy density 
${\cal E} = -p + \mu\rho + {1\over 2}\nu\rho_3$:
\begin{eqnarray}
p & = &{1\over 3\pi^2}\int_{0}^{k_{\rm Fp}}dk {k^4\over(k^2+M^*{}^2)^{1/2}}
      +{1\over 3\pi^2}\int_{0}^{k_{\rm Fn}}dk {k^4\over(k^2+M^*{}^2)^{1/2}}
      \nonumber\\
 & &\null + {1\over 2c_{\rm v}^2}\, W^2
          +{1\over 2c_{\rho}^2}\, R^2
          -{1\over 2c_{\rm s}^2}\, \Phi^2
          +\sum_{i,j,k}{\bar a}_{ijk}\,\Phi^i\, W^{2j}\, R^{2k}
         \ ,  \label{eq:press}\\[4pt]
{\cal E} &=&{1\over \pi^2}\int_{0}^{k_{\rm Fp}}dk\, k^2(k^2+M^*{}^2)^{1/2}
      +{1\over \pi^2}\int_{0}^{k_{\rm Fn}}dk\, k^2(k^2+M^*{}^2)^{1/2}
      \nonumber\\
 & & \null + W \rho 
    + {1\over 2}\, R \rho_3
    - {1\over 2c_{\rm v}^2}\, W^2
    - {1\over 2c_{\rho}^2}\, R^2
  + {1\over 2c_{\rm s}^2}\, \Phi^2 
   -\sum_{i,j,k}{\bar a}_{ijk}\,\Phi^i\, W^{2j}\, R^{2k}
   \ .        \label{eq:energy}
\end{eqnarray}
Here the ratios $c^2_i=g^2_i/m^2_i$ and
${\bar a}_{ijk}\equiv a_{ijk}/(g_s^i g_{\rm v}^{2j} g_{\rho}^{2k})$ have
been introduced for convenience.

As noted earlier, the pressure in Eq.~(\ref{eq:press}) and the energy 
density in
Eq.~(\ref{eq:energy}) also contain vacuum contributions arising from the
partition function of the nucleons. 
However, at least at the one-baryon-loop level, these vacuum terms can be
absorbed in the definition of the nonlinear couplings in 
Eq.~(\ref{eq:Pot2}) \cite{FTS95}, and thus we include explicitly only
the contributions from valence nucleons.

At zero temperature, the stationarity conditions of Eq.~(\ref{eq:extremum}) 
with fixed chemical potentials are equivalent to an extremization of
the energy at fixed baryon and isovector density.
This leads to the self-consistency equations
\begin{eqnarray}
{1\over c_{\rm s}^2}\, \Phi
  -\sum_{i,j,k} i\, {\bar a}_{ijk}\, \Phi^{i-1}\, W^{2j}\, R^{2k}
           &=& \rho_{\rm s} \ ,\label{eq:scalar}\\
{1\over c_{\rm v}^2}\, W
  +\sum_{i,j,k} 2j\, {\bar a}_{ijk}\, \Phi^i\, W^{2j-1}\, R^{2k}
        &=& \rho\ , \label{eq:vector}\\
{1\over c_{\rho}^2}\, R
  +\sum_{i,j,k} 2k\, {\bar a}_{ijk}\, \Phi^i\, W^{2j}\, R^{2k-1}
        &=& {1\over 2}\rho_3\ , \label{eq:Rrho}
\end{eqnarray}
where the scalar density is given by
\begin{eqnarray}
\rho_{\rm s} = {M^*\over \pi^2}
\int_{0}^{k_{\rm Fp}}dk {k^2\over(k^2+M^*{}^2)^{1/2}}
      +{M^*\over \pi^2}\int_{0}^{k_{\rm Fn}}dk {k^2\over(k^2+M^*{}^2)^{1/2}}
           \ . \label{eq:rhos}
\end{eqnarray}

Because of the factor of $1/2$ on the right-hand side of Eq.~(\ref{eq:Rrho}),
it follows that the relevant expansion parameters for the energy density
are $\Phi / M$, $W/M$, and $2R/M$.
Moreover, by dividing $\cal E$ by $M^4$ and expressing the result in
terms of these expansion parameters, one can identify the scaled couplings
that should all be of roughly the same size if they are ``natural'',
namely,
$$ {1\over 2 c^2_{\rm s}M^2}\ , \quad
   {1\over 2 c^2_{\rm v}M^2}\ , \quad
   {1\over 8 c^2_{\rho}M^2}\ , \quad {\rm and} \quad
   {{\bar a}_{ijk} M^{i+2j+2k-4}\over 2^{2k}}\ .
      \label{eq:natural}
$$

%

%
\section{The High-Density Limit}
The coupling constants
$c^2_{\rm v}$, $c^2_{\rm s}$, $c^2_{\rho}$, and 
${\bar a}_{ijk}$ in Eqs.~(\ref{eq:press}) and (\ref{eq:energy})
enter as unknown model parameters.
According to the generally accepted procedure, these parameters will be
chosen to reproduce the properties of nuclear matter near equilibrium.
The basic ingredient in many astrophysical problems, {\em e.g.},
neutron-star calculations, is the EOS
\begin{eqnarray}
p = p({\cal E}) \ , \label{eq:EOS}
\end{eqnarray}
which is then extrapolated into the neutron-rich and high-density regime.
Anticipating the results of the next section,
one can expect that different parameter sets that lead to
identical equilibrium properties produce qualitatively similar equations of
state at low densities. 
The relevant question is whether this qualitatively similar behavior
persists at high densities, particularly in the regime important
for neutron stars.
As a first step in this direction, we investigate the high-density limit of
the EOS generated by the model introduced in the previous section.

To make the discussion more transparent, we focus here on pure neutron matter
($k_{\rm Fp}=0, k_{\rm Fn}\equiv k_{\rm F}$), although nuclear matter in 
$\beta$-decay equilibrium
with a finite proton to neutron ratio is necessary to achieve accurate
results for maximum neutron star masses.
We will return to this issue in the next section.

It is clear that a sufficiently large number of couplings introduces
a high degree of flexibility. 
Due to the nonlinearity of the problem, not all
families of parameter sets lead to physically acceptable results, which
provides one way to restrict the parameter space.
Classes of models can be ruled out if basic physical 
requirements are violated.
For example, one certainly requires that the pressure $p$ be a smooth 
function of the energy density ${\cal E}$.
Moreover, it is necessary that the speed of (first) sound $c_1$ respect 
causality and also be real, to ensure stability.
That is,
$$ 0 \leq c_1^2={\partial p\over \partial {\cal E}}\leq 1 \, .$$
In addition to these general principles, we require
a positive and bounded value of the nucleon effective mass, {\em i.e.},
\begin{eqnarray}
0\,\leq\,M^*\,\leq\,M\ .\label{eq:restoration}
\end{eqnarray}
This is motivated by the expectation that physically reasonable models
will demonstrate some degree of chiral-symmetry restoration at finite
density, leading to a reduction in the nucleon mass, with the most
extreme situation corresponding to total restoration of the symmetry.
The consequences of Eq.~(\ref{eq:restoration}) will become more transparent
in the following.

By specializing the formalism of the preceding section to pure 
neutron matter and by
using the self-consistency equations (\ref{eq:scalar})--(\ref{eq:Rrho}),
the pressure and energy density can be expressed as
\begin{eqnarray}
p & \equiv &p_0(k_{\rm F},M^*)+\Delta p(\Phi,W,R)\nonumber\\
  & = &\null {1\over 3\pi^2}\int_{0}^{k_{\rm F}}dk 
     {k^4\over(k^2+M^*{}^2)^{1/2}}
      \nonumber\\
 & &\null + {1\over 2c_{\rm v}^2}\, W^2
          +{1\over 2c_{\rho}^2}\, R^2
          -{1\over 2c_{\rm s}^2}\, \Phi^2
          +\sum_{i,j,k}{\bar a}_{ijk}\,\Phi^i\, W^{2j}\, R^{2k} \ ,
           \label{eq:pressn}\\[4pt]
{\cal E} &\equiv&{\cal E}_0(k_{\rm F},M^*)+\Delta 
           {\cal E}(\Phi,W,R)\nonumber\\
 & = &\null{1\over \pi^2}\int_{0}^{k_{\rm F}}dk\, k^2(k^2+M^*{}^2)^{1/2}
      \nonumber\\
 & & \null +{1\over 2c_{\rm v}^2}\, W^2
    +{1\over 2c_{\rho}^2}\, R^2
  + {1\over 2c_{\rm s}^2}\, \Phi^2
  +\sum_{i,j,k}(2j+2k-1)\,{\bar a}_{ijk}\,\Phi^i\, W^{2j}\, R^{2k} \ ,
           \label{eq:energyn}
\end{eqnarray}
where $p_0$ and ${\cal E}_0$ denote the results for a 
relativistic, noninteracting gas of spin-1/2 baryons with mass $M^*$.
Note that these expressions include, as a special case, models where only
the ${\bar a}_{0jk}$ are nonzero, so that there are no scalar--vector
couplings.
Moreover, the isovector density in Eq.~(\ref{eq:Rrho}) is replaced by
\begin{eqnarray} 
\rho_3=-\rho=-{1\over 3\pi^2}k_{\rm F}^3 \ . 
\end{eqnarray}

At high densities, the left-hand sides of the
self-consistency equations (\ref{eq:vector})
and (\ref{eq:Rrho}) must grow linearly in $\rho$, and thus
we start with the {\em Ansatz}
\begin{eqnarray}
\lim_{\rho \to \infty}W =  w_0 \rho^{\alpha}
\quad , \quad
\lim_{\rho \to \infty}R = r_0 \rho^{\beta} \ ,
\end{eqnarray}
where $0<\alpha , \beta \leq 1$, $w_0 >0$, and $r_0 <0$.
Since we assume that the effective mass is bounded, we can replace
the scalar field by the limit
\begin{eqnarray}
\Phi_{\infty} \equiv \lim_{\rho \to \infty} (M-M^*) \leq M \ .
\label{eq:limitphi}
\end{eqnarray}
To fulfill the resulting self-consistency equations:
\begin{eqnarray}
{1\over c_{\rm v}^2}\, W
  +\sum_{i,j,k} 2j\, {\bar a}_{ijk}\, \Phi^i_{\infty}\, W^{2j-1}\, R^{2k}
        &=& \rho\ , \label{eq:Wlimit}\\
{1\over c_{\rho}^2}\, R
  +\sum_{i,j,k} 2k\, {\bar a}_{ijk}\, \Phi^i_{\infty}\, W^{2j}\, R^{2k-1}
        &=& {1\over 2}\rho_3 = -{1\over 2}\rho
        \ , \label{eq:Rlimit}
\end{eqnarray}
there must be integers $(j_m,k_m)$ and $(j^{\prime}_m,k^{\prime}_m)$
with
\begin{eqnarray}
(2j_m-1)\alpha + 2k_m\beta = 1 
   \,&>&\, (2j-1)\alpha + 2k\beta
 \quad\hbox{for all}\quad j\neq j_m \, ,\, k \neq k_m \ ,
   \label{eq:Winteger}\\
2j^{\prime}_m \alpha + (2k^{\prime}_m - 1)\beta = 1 
   \,&>&\, 2j\alpha + (2k - 1)\beta
   \quad\hbox{for all}\quad j\neq j^{\prime}_m \, ,\, k \neq k^{\prime}_m\ .
   \label{eq:Rinteger}
\end{eqnarray}
(This assumes that only one term in each sum produces the leading
asymptotic behavior; if this actually happens for more than one term
in a sum, the conclusions below are unchanged.)
Using
\begin{eqnarray}
\lim_{\rho \to \infty}p_0  = {(3\pi^2)^{1/3}\over 4}\rho^{4/3}
	   +O(\rho^{2/3})
\quad , \quad
\lim_{\rho \to \infty}{\cal E}_0  =
            {3(3\pi^2)^{1/3}\over 4}\rho^{4/3}+O(\rho^{2/3})\ ,
\end{eqnarray}
the leading contributions to the pressure and the energy density are 
found to be
\begin{eqnarray}
& & \lim_{\rho \to \infty}p = {(3\pi^2)^{1/3}\over 4}\rho^{4/3}
    + {1\over 2c_{\rm v}^2}w_0^2 \rho^{2\alpha}
    + {1\over 2c_{\rho}^2}r_0^2 \rho^{2\beta}  \nonumber\\
& &\null \qquad\qquad
   +w_0^{2j_m}  r_0^{2k_m} \rho^{1+\alpha}
   \sum_{i} {\bar a}_{ij_m k_m} \Phi_{\infty}^i
         +w_0^{2j^{\prime}_m}  r_0^{2k^{\prime}_m} \rho^{1+\beta}
   \sum_{i} {\bar a}_{ij^{\prime}_m k^{\prime}_m} \Phi_{\infty}^i \ ,
   \label{eq:plimit}\\
& & \lim_{\rho \to \infty}{\cal E} = {3(3\pi^2)^{1/3}\over 4}\rho^{4/3}
    + {1\over 2c_{\rm v}^2}w_0^2 \rho^{2\alpha}
    + {1\over 2c_{\rho}^2}r_0^2 \rho^{2\beta}  \nonumber\\
& &\null \qquad\qquad
   +w_0^{2j_m}  r_0^{2k_m} \rho^{1+\alpha}(2j_m+2k_m-1)
   \sum_{i} {\bar a}_{ij_m k_m} \Phi_{\infty}^i\nonumber\\
& &\null \qquad\qquad
    +w_0^{2j^{\prime}_m}  r_0^{2k^{\prime}_m} \rho^{1+\beta}
	 (2j^{\prime}_m+2k^{\prime}_m-1)
   \sum_{i} {\bar a}_{ij^{\prime}_m k^{\prime}_m} \Phi_{\infty}^i \ .
   \label{eq:elimit}
\end{eqnarray}

To this point, the discussion is rather general.
To make the conclusions more concrete, we discuss two distinct situations 
separately:

\begin{enumerate}
\item No coupling between $W$ and $R$.

In this special case, the asymptotic behavior
of the fields is governed by their
highest powers in the potential (\ref{eq:Pot2}).
From Eqs.~(\ref{eq:Winteger}) and (\ref{eq:Rinteger}), we obtain
$$ \alpha={1\over 2j_{\rm max} -1} \quad , \quad 
   \beta={1\over 2k_{\rm max}-1} \ . $$
For $j_{\rm max}=1\, , k_{\rm max}\geq 1$ or
$j_{\rm max}\geq 1\, , k_{\rm max}=1$, the quadratic terms dominate the
right-hand sides of Eqs.~(\ref{eq:plimit}) and (\ref{eq:elimit}), 
so that
$$
     \lim_{\rho \to \infty} {1\over c_{\rm v}^2}W^2
   \propto \rho^2 \quad \hbox{or} \quad
     \lim_{\rho \to \infty} {1\over c_{\rho}^2}R^2 
   \propto \rho^2 \ .
$$
This case includes the original version of the Walecka model
\cite{WALECKA74} and generates the limiting behavior
\begin{eqnarray}
& & \lim_{\rho \to \infty}p = {\cal E} \ .
   \label{eq:limit1}
\end{eqnarray}
The sums in $\Delta p$ and $\Delta {\cal E}$
contribute to the leading order only if $j_{\rm max}=2\, ,\, 
k_{\rm max}\geq 2$
or $j_{\rm max}\geq 2\, ,\, k_{\rm max}=2$. 
In this case, the quadratic terms can
be neglected, and $\Delta p$ and $\Delta{\cal E}$ are of 
the same order as the contributions from the ideal-Fermi-gas terms. 
However, the factors in Eq.~(\ref{eq:elimit}) conspire such that
\begin{eqnarray}
\lim_{\rho \to \infty} \Delta p = {1\over 3}\Delta{\cal E} \ ,
\label{eq:arrange}
\end{eqnarray}
and the functional form of the limiting EOS resembles that of an
ideal Fermi gas:
\begin{eqnarray}
\lim_{\rho \to \infty} p = {1\over 3}{\cal E} \ .
   \label{eq:limit2}
\end{eqnarray}
In the remaining cases ($j_{\rm max}>2 , k_{\rm max}>2$), the 
dominant contributions arise
solely from $p_0$ and ${\cal E}_0$, which also leads to 
Eq.~(\ref{eq:limit2}).
Note here the importance of Eq.~(\ref{eq:restoration}), which implies 
that in the high-density limit, $M^*$ becomes negligible, at least 
to leading order.
\item At least one coupling between $W$ and $R$.

From Eqs~(\ref{eq:Winteger}) and (\ref{eq:Rinteger}),
 it follows directly that
$$ \alpha\leq{1\over 3} \quad\hbox{and}\quad
    \beta\leq{1\over 3} \ .
$$
The quadratic contributions in the fields are negligible, and $\Delta p$ and 
$\Delta {\cal E}$ contribute to the leading term only
if $\alpha=\beta=1/3$, where again Eq.~(\ref{eq:arrange}) holds. 
In any event, this leads to the limit of Eq.~(\ref{eq:limit2}).
\end{enumerate}

To summarize, we conclude that the high-density limit of the EOS is strongly
influenced by nonlinear meson--meson interactions, which agrees with
the conclusion of Bodmer and Price \cite{BODMER89,BODMER91}. 
The limit in Eq.~(\ref{eq:limit1}) obtained in the original version 
of the Walecka model \cite{WALECKA74} is 
a special case; in the more general situation, the
nuclear matter EOS approaches that of an ideal Fermi gas,
given by Eq.~(\ref{eq:limit2}).
We will show in the next section that these two limits can be achieved
using different models with parameter sets that reproduce
the same equilibrium properties of nuclear matter.

%

%
\section{Consequences for Neutron Stars}
The high-density limit of the EOS and the way in which the matter
approaches the asymptotic regime have important consequences
in neutron star calculations.
The masses and radii of stars are sensitive to the stiffness of the EOS, thus
providing a quantitative measure for studying the impact of the nonlinear
interaction terms in Eq.~(\ref{eq:Pot2}).

To be specific, it is necessary to choose an explicit potential, and we
begin with the form
\begin{eqnarray}
\sum_{i,j,k}a_{ijk}\,\phi^i\, (V_{\mu}V^{\mu})^j\,
			   ({\bf b_{\mu}} \bolddot{\bf b}^{\mu})^k
   = -{\kappa\over 3!}\phi^3-{\lambda\over 4!}\phi^4
	  +{\zeta\over 4!}g_{\rm v}^4(V_{\mu}V^{\mu})^2
  +{\xi\over 4!}g_{\rho}^4({\bf b_{\mu}} \bolddot{\bf b}^{\mu})^2\ ,
\label{eq:poly4}
\end{eqnarray}
which includes a subset of the
meson self-interactions up to fourth order in the fields.
As discussed in the Introduction, setting some of the allowed
cubic and quartic
couplings to zero is ``unnatural'', but as we will discover, the model
defined by Eq.~(\ref{eq:poly4}) is already general enough to produce
significant differences in predicted neutron star masses, and restoring
the omitted couplings will lead to even more variation in the results.
Moreover, the present model can be related to the most common models 
discussed in the literature, and it generalizes them to include a
nonlinear isovector interaction.
The motivation for adding the quartic rho-meson term is that one expects
this coupling to be essentially unconstrained by normal nuclear observables,
where the neutron--proton asymmetry is low, but it may have significant
impact on the neutron-rich matter in neutron stars.
As noted earlier, since the meson fields are effective (Kohn--Sham)
potentials, we are not concerned here with their elementary excitations, and
considerations of causality \cite{VELO69,VELO73} are unimportant.

In nuclear matter calculations, this model contains seven free parameters.
The polynomial in Eq.~(\ref{eq:poly4}) contains four couplings that we
may write as ${\bar\kappa}\equiv \kappa / g_{\rm s}^3$,
${\bar\lambda}\equiv \lambda /g_{\rm s}^4$,
$\zeta$, and $\xi$; in addition,
values for the three ratios $c_i^2=g^2_i/m^2_i$ ($i = $ s, v, $\rho$)
are needed.
Five of the seven parameters can be chosen to reproduce the
equilibrium properties of symmetric nuclear matter, which we take as
the equilibrium density and binding energy ($\rho_0$, $-e_0$), 
the nucleon effective (or Dirac) mass at equilibrium ($M^\ast_0$), 
the compression modulus ($K_0$), and the bulk symmetry energy ($a_4$).
The first three of these are tightly constrained \cite{FURNSTAHL93}, 
whereas the latter two are not.
In principle, the sensitivity of the high-density EOS to reasonable 
variations in $K_0$ and $a_4$ could be examined, but for simplicity, we
keep their values fixed in most of our calculations.
The ``standard'' set of equilibrium properties used here are listed
in Table~\ref{tab:one}; these are motivated by successful descriptions of
bulk and single-particle nuclear properties \cite{FURNSTAHL93,FTS95,FST96}.
The nucleon mass is fixed at its empirical value 
($M = 939\,{\rm MeV}$).

Our primary goal is to study the influence of the nonlinear vector-meson
interactions on neutron star masses.
Since there are more free couplings than normalization conditions, we
proceed as follows:
We choose values for the couplings $\zeta$ and $\xi$
and determine the remaining couplings by requiring that they reproduce
the desired equilibrium properties. 
This is achieved by solving a set of transcendental equations
that relate the parameters directly to the nuclear matter 
properties \cite{BODMER91,FST96}.
Although we have no specific guidance on the allowed values of $\zeta$ and 
$\xi$, we rely on the assumption of naturalness, and based on the
discussion at the end of Sec.~II, we observe that the following parameter
combinations should all be of roughly equal size:
$$ {1\over 2 c^2_{\rm s}M^2}\ , \quad
   {1\over 2 c^2_{\rm v}M^2}\ , \quad
   {1\over 8 c^2_{\rho}M^2}\ , \quad
   {{\bar\kappa} \over 6 M}\ , \quad
   {{\bar\lambda} \over 24}\ , \quad
   {\zeta \over 24}\ , \quad {\rm and} \quad
   {\xi \over 384}\ .
\label{eq:polynatural}
$$
Typical values for the first three parameters are between 0.001 and
0.002, so that the natural values of $\zeta$ and $\xi$ are roughly 
limited to $0\leq\zeta\alt 0.06$ and $0\leq\xi\alt 1.0$.
(To avoid abnormal solutions of the vector field equations, 
{\em i.e.}, those with finite mean fields at zero density,
$\zeta$ and $\xi$ must be positive.\footnote{%
Note that positive $\zeta$ and $\xi$ imply that the resulting nonlinear 
interactions are {\em attractive}.
This constraint on the highest-order vector interactions appears to be 
general and explains why our earlier analysis finds that nonlinear 
interactions soften the equation of state.})
We will include results for vanishing $\zeta$ and $\xi$, which are
in a strict sense unnatural, in order to make contact with 
earlier calculations.
For a more thorough discussion of naturalness and its implications,
see Ref.~\cite{FST96}.

Using the notation of Sec.~II, the self-consistency equations 
~(\ref{eq:scalar})--(\ref{eq:Rrho}) can be written as
\begin{eqnarray}
{1\over c_{\rm s}^2}\, \Phi
+{\bar\kappa\over 2}\Phi^2+{\bar\lambda\over 6}\Phi^3
           &=& \rho_{\rm s} \ ,\label{eq:scalar4}\\
W\left({1\over c_{\rm v}^2}+{\zeta\over 6}W^2\right)
        &=& \rho\ , \label{eq:vector4}\\
R\left({1\over c_{\rho}^2}+{\xi\over 6}R^2\right)
        &=& {1\over 2}\rho_3 =-{1\over 2}\rho \ . \label{eq:Rrho4}
\end{eqnarray}
The expressions for the pressure and the energy density follow as
\begin{eqnarray}
p & = &{1\over 3\pi^2}\int_{0}^{k_{\rm F}}dk {k^4\over(k^2+M^*{}^2)^{1/2}}
      \nonumber\\
 & &\null + {1\over 2c_{\rm v}^2}\, W^2 + {\zeta\over 24}\,W^4
          +{1\over 2c_{\rho}^2}\, R^2 + {\xi\over 24}\,R^4
          -{1\over 2c_{\rm s}^2}\, \Phi^2-{\bar\kappa\over 6}\Phi^3
	  -{\bar\lambda\over 24}\Phi^4 \ ,
           \label{eq:press4}\\
{\cal E} &=&{1\over \pi^2}\int_{0}^{k_{\rm F}}dk\, k^2(k^2+M^*{}^2)^{1/2}
       \nonumber\\
 & &\null + {1\over 2c_{\rm v}^2}\, W^2 + {\zeta\over 8}\,W^4
          +{1\over 2c_{\rho}^2}\, R^2 + {\xi\over 8}\,R^4
          +{1\over 2c_{\rm s}^2}\, \Phi^2+{\bar\kappa\over 6}\Phi^3
	  +{\bar\lambda\over 24}\Phi^4 \ .
           \label{eq:energy4}
\end{eqnarray}

We begin our discussion with the model introduced by Bodmer and Price
\cite{BODMER89}, which corresponds to $\xi=0$. 
According to the discussion in the preceding section, this model has
the interesting feature that the high-density EOS of pure neutron matter
approaches $ p={\cal E}$, while in symmetric matter, 
where the mean-field of the $\rho$ meson
vanishes, the EOS approaches the massless Fermi gas limit, given by
Eq.~(\ref{eq:limit2}). 
In Fig.~\ref{fig:f1}, we show the binding energy 
curves for symmetric and pure neutron matter for different values of the
nonlinear coupling $\zeta$. 
We emphasize that all parametrizations reproduce the same equilibrium 
properties listed in Table~\ref{tab:one}.

At low densities, all the curves approach a common limit, because
the nonlinear terms do not contribute at leading order in a
low-density expansion. 
At higher densities, the softening of the EOS as $\zeta$ increases is
clearly visible, at least for symmetric matter. 
The softening in neutron matter is more apparent in Fig.~\ref{fig:f2}.
In the regime of intermediate density, 
$200\alt {\cal E}\alt 1000\,{\rm MeV}/{\rm fm}^{3}$,
the EOS becomes softer with increasing values of $\zeta$ \cite{BODMER91}.
To study the approach to the asymptotic limit, we examine the nonleading 
terms in the high-density expansion:
\begin{eqnarray}
& & {\lim_{{\cal E}\to \infty}}\ p = {\cal E} 
      -{1\over 6\pi^2}\left(
      {72\pi^4\over 4c_{\rm v}^2 + c_{\rho}^2}\right)^{2/3}{\cal E}^{2/3}
      + O({\cal E}^{1/3})
      \quad\hbox{for}\quad\zeta=0 \ ,\label{eq:asymp1a}\\
& & {\lim_{{\cal E}\to \infty}}\ p = {\cal E} 
      -{1\over 6\pi^2}\left(
      {72\pi^4\over  c_{\rho}^2}\right)^{2/3}
      \left[1+\Bigr({2\over \pi^2\zeta}\Bigl)^{1/3}\right] {\cal E}^{2/3}
      + O({\cal E}^{1/3})
      \quad\hbox{for}\quad\zeta\neq0 \ ,\label{eq:asymp1b}
\end{eqnarray}
which reveals two important features.
First, the $\zeta \neq 0$ results are {\em nonanalytic\/} in $\zeta$.
Thus one cannot reproduce Eq.~(\ref{eq:asymp1a}) by taking
the $\zeta\rightarrow 0$ limit of Eq.~(\ref{eq:asymp1b});
at least as far as $\zeta$ is concerned,
the high-density expansion is essentially a strong coupling expansion.
Second, we observe that the coefficient of the nonleading term is smaller
for $\zeta = 0$ for two reasons: the appearance of the isoscalar coupling
$c_{\rm v}^2$ in the denominator and the absence of the multiplicative
factor containing $\zeta$.
(Note that $c_{\rho}^2$ is independent of $\zeta$.)
These two features 
produce a coefficient that is roughly an order of magnitude
smaller for $\zeta = 0$ than for $\zeta \not= 0$, which explains the
relatively slow approach to the asymptotic limit in the latter case, as
is evident from Fig.~\ref{fig:f2}.

The consequences for neutron stars can be studied in Fig.~\ref{fig:f3},
where the star masses are shown as a function of the central mass
density $\rho_c$.
As expected from Fig.~\ref{fig:f2},
the maximum mass decreases with increasing $\zeta$.
This decrease is substantial:
from $M_{\rm max}=2.9\,M_{\odot}$ for $\zeta=0$ to
$M_{\rm max}=2.1\,M_{\odot}$ for $\zeta=0.06$, which is roughly 30\%.
The shifts in the maximum mass are most dramatic for small couplings;
for larger couplings, the softening effects begin to saturate.

To understand this result, it is useful to identify the regime of energy
density that is most important in determining the mass of the star.
This regime can be deduced from Fig.~\ref{fig:f4}, where we show the
radial mass density distributions for several neutron stars, as well as
the corresponding energy densities.
Observe that most of the mass is generated at radii between 6 and 12 km,
which corresponds to energy densities of several hundred MeV/fm${}^3$.
As can be seen from Fig.~\ref{fig:f2}, this includes the regime where the
EOS is sensitive to $\zeta$, because this is where the $W^4$
contribution to the EOS begins to become important.
Note, however, that the contribution of the $W^4$ term in this regime is
still smaller than that of the $W^2$ term; the quartic term does not begin
to dominate until the energy density reaches several thousand MeV/fm${}^3$,
as indicated by the coalescence of the dashed and dotted curves in
Fig.~\ref{fig:f2}.

We now return to the general form in Eq.~(\ref{eq:poly4}) 
and allow nonlinear self-interactions of the $\rho$ mesons. 
In contrast to the isoscalar coupling $\zeta$, the new quantity 
$\xi$ does not enter in the calculation of our five ``standard''
equilibrium properties of nuclear matter,
and the other parameters are determined independently.
This is true even for the symmetry energy, because the new coupling $\xi$
first appears at order $(N-Z)^4$ in an
expansion around symmetric nuclear matter. 
In principle, $\xi$ could be constrained by
fits to liquid-drop expansions of the energy, but in most such
fits this parameter is set to zero.
(We found only one nonzero value in the literature \cite{MYERS95}.)
Thus, at present, contributions to the symmetry energy beyond terms of 
order $(N-Z)^2$ are practically unconstrained.

Fig.~\ref{fig:f5} shows the binding energy of nuclear matter as a function
of the proton fraction $y$ for two different densities.
The curves are calculated for various values of $\xi$
with $\zeta$ held fixed.
For clarity, we plot the fractional shift in $\cal E$ relative to its
value with $\xi = 0$.
At normal nuclear density, results for different $\xi$ are virtually
indistinguishable, but at high density, the curves differ by a few percent 
when the proton fraction becomes very small.
This demonstrates that it is possible to generate families of models
that reproduce identical properties of nuclear matter at low and normal
densities, but which generate different predictions at high densities.

According to the analysis in the preceding section, the EOS 
asymptotically approaches the massless Fermi gas limit for both 
symmetric and neutron matter.
The high-density expansion for neutron matter corresponding to 
Eq.~(\ref{eq:asymp1b}) is
\begin{eqnarray}
 {\lim_{{\cal E}\to \infty}}\ p = {1\over 3}{\cal E} 
      +{2\pi\over 3}{\textstyle 
      \left[{\textstyle 1\over 
          \textstyle\vphantom{C_{\rm v}^{2^2}}c_{\rm v}^2}
      \Bigr({\textstyle 2\over 
          \textstyle\vphantom{C_{\rm v}^{2^2}} \pi^2\zeta}\Bigl)^{2/3}
           +{\textstyle 1\over 
          \textstyle\vphantom{C_{\rm v}^{2^2}} c_{\rho}^2}\Bigr(
      {\textstyle 1\over 
           \textstyle\vphantom{C_{\rm v}^{2^2}} \pi^2\xi}\Bigl)^{2/3}\right]
	   \over\textstyle 
      \left[1+\Bigr({\textstyle 2\over 
           \textstyle\vphantom{C_{\rm v}^{2^2}} \pi^2\zeta}\Bigl)^{1/3}
           +\Bigr({\textstyle 1\over 
            \textstyle\vphantom{C_{\rm v}^{2^2}} \pi^2\xi}\Bigl)^{1/3}
        \right]^{1/2}}\, {\cal E}^{1/2}
      + O({\cal E}^{0}) 
       \ .\label{eq:asymp2}
\end{eqnarray}
In contrast to the previous case, the asymptotic 
limit is now approached from above. 
Moreover, the analytical form has changed, since Eq.~(\ref{eq:asymp2})
indicates a series in powers of ${\cal E}^{1/2}$.

The dependence of the neutron star masses on $\xi$ follows the same trend
as found for $\zeta$.
This can be gleaned from Fig.~\ref{fig:f6}, which shows results for pure
neutron matter.
For fixed $\zeta$, the maximum mass decreases with increasing $\xi$, 
although the effect of this new parameter is smaller (less than a 10\%
change in the maximum mass), given the limitation
on parameter values imposed by naturalness.

A more complete picture of $M_{\rm max}$ is given in Fig.~\ref{fig:f7}, 
where the variations with both $\zeta$ and $\xi$ are shown.
It is apparent that parameter sets that yield identical properties near
nuclear equilibrium can still generate values of the maximum mass that 
differ by as much as one solar mass.
We find variations between $M_{\rm max} = 2.9\, M_{\odot}$
and $M_{\rm max} = 1.9\, M_{\odot}$ for stars composed of pure neutron
matter.
The masses are most sensitive to changes at small values of the couplings, 
particularly for $\zeta \approx 0$, which can be related to the 
nonanalytic form of the EOS in terms of $\zeta$ and $\xi$.
[See Eqs.~(\ref{eq:asymp1a})--(\ref{eq:asymp2}).]
As seen earlier, the effect of $\xi$ is smaller than that of $\zeta$.

So far, we have considered pure neutron matter, which gives
only a qualitative picture of neutron star properties.
For a more realistic description, it is necessary to consider beta-stable
matter, as this includes the protons, which softens the EOS.
Maximum neutron star masses for beta-stable matter are also
shown in Fig.~\ref{fig:f7}. 
The dependence on the isoscalar coupling $\zeta$ is similar to that obtained
earlier (the maximum mass varies between $M_{\rm max} = 2.8\, M_{\odot}$
and $M_{\rm max} = 1.8\, M_{\odot}$), 
but the influence of the isovector coupling $\xi$ is less drastic,
since the matter becomes significantly more symmetric in the region
that gives the largest contribution to the mass.

In Fig.~\ref{fig:f8}, we examine the dependence of the maximum mass on the
compression modulus at equilibrium, which is not particularly well known.
(Relativistic mean-field models with $200 \alt K_0 \alt 350\, {\rm MeV}$
can produce accurate nuclear binding-energy systematics and surface
energetics \cite{FST96}.)
The shaded band shows the total predicted variation in maximum mass when 
both $\zeta$ and $\xi$ are varied within the bounds imposed by naturalness.
The dashed curve shows the predicted variation when $K_0$ is varied at
fixed $\zeta$ and $\xi$.
Evidently, the variations in the maximum mass arising from the possibility
of nonlinear vector meson interactions is much greater than that arising
from the uncertainty in the compressibility.

These results raise the interesting question of whether the only significant
nonlinearity is the quartic, isoscalar vector interaction.
In other words, once one takes $\zeta\not= 0$ to soften the EOS at high
densities, does the addition of further nonlinearities produce only small
effects?
To examine this question, we extend the model of Eq.~(\ref{eq:poly4})
to include sixth-order and eighth-order terms involving the 
vector-isoscalar meson:
\begin{eqnarray}
\Delta {\cal V}' =
   -{\zeta'\over 6!}g_{\rm v}^6(V_{\mu}V^{\mu})^3 
  - {\zeta''\over 8!}g_{\rm v}^8(V_{\mu}V^{\mu})^4 \ .
\label{eq:poly8}
\end{eqnarray}
Thus $\zeta'/6!$ and $\zeta''/8!$ are the relevant ratios to be included in 
the parameter list given earlier, and we will initially set $\xi$ to zero 
and examine the consequences of varying $\zeta$, $\zeta'$, and $\zeta''$ 
within the bounds imposed by naturalness.

Figure \ref{fig:f9} shows neutron star masses as higher-order vector
nonlinearities are included sequentially.
Evidently, the quartic interactions are the most important, producing a
roughly 30\% variation in the maximum.
The effects of the sixth-order term are quite modest (roughly 10\%), while
the eighth-order contributions are essentially negligible (roughly 2\%).
Here the parameters are varied within the natural ranges
$0 \leq \zeta \leq 0.06$, 
$0 \leq \zeta' \leq 1.2$, and
$0 \leq \zeta'' \leq 60$.
Thus we have the encouraging result that once the $W^4$ interaction has been
accurately calibrated, contributions from higher-order nonlinearities are
relatively unimportant.
To indicate the most extreme reduction in maximum mass possible in
the present model, the cross in Fig.~\ref{fig:f9} shows $M_{\rm max} =
1.58\, M_\odot$, which is obtained for {\em beta-stable matter\/} when the 
couplings $\zeta = 0.06$, $\zeta' = 1.2$, $\zeta'' = 60$, and $\xi = 1.5$ are
included.
Note that this value of $M_{\rm max}$ is only slightly ($\approx 10$\%)
larger than that of the most massive observed neutron stars.

Our analysis to this point has revealed significant model and parameter
dependence in the high-density EOS.
It is therefore of interest to see if these variations influence predictions 
arising from other relevant dynamics in systems with high densities.
As an example, we study the effect of the high-density hadronic EOS on
the existence of quark-matter cores in neutron stars.
We adopt a simple two-phase model \cite{BAYM76,SW86} based on a
first-order (van der Waals) phase transition between the hadronic and
quark phase.
Although there are indications from QCD lattice calculations that
the hadron/quark phase transition is second-order at vanishing chemical
potential \cite{BROWN90,CHRIST92} (for massless quarks), 
the true behavior of the transition
at finite density (and indeed, whether one actually exists) is unknown at
present.
Until more reliable information is available, one must resort to
a separate description of the quark phase and the hadronic phase.
Moreover, whereas a detailed description involving beta-stable matter 
requires a careful treatment of the phase transition in systems with
two conserved charges (baryon number and isospin) \cite{GLENN92,MULLER95},
here we will be satisfied with a qualitative discussion based on pure
neutron matter.
This is certainly reasonable, given the large uncertainties we have
already found in the hadronic EOS at high density.

We adopt the simple EOS involving massless $u$ and $d$ quarks given by
\begin{eqnarray}
 p ={1\over 3}{\cal E}-{4\over 3}b
       \ ,\label{eq:bageos}
\end{eqnarray}
where the confinement property of QCD (or alternatively, the anomaly in
the trace of the energy-momentum tensor) is modeled by a positive constant
$b$, which represents the energy per unit volume in the vacuum.

We return to the hadronic model of Eq.~(\ref{eq:poly4}), which leads to
the high-density expansion in Eq.~(\ref{eq:asymp2}).
Our discussion is based on the simple observation that
independent of the actual nature of the transition, it is driven purely
by the energetics in the two phases.

By comparing Eq.~(\ref{eq:bageos}) with the expansion in
Eq.~(\ref{eq:asymp2}), one observes that in the quark phase, 
the limit $p={\cal E}/3$ is approached from below,
whereas in neutron matter, for the general case $\zeta > 0$ and $\xi > 0$,
the limit is approached from above. 
To decide whether a transition takes place, it is necessary to
compare the energy/baryon in both phases.  
For the quark phase it is expressed as \cite{BAYM76}
\begin{eqnarray}
 {\cal E}/\rho = {3\over 4}\pi^{2/3}
			f(\alpha_s) \rho^{1/3} + b/ \rho
       \ ,\label{eq:bageb}
\end{eqnarray}
with
$$ f(\alpha_s)=(1+2^{4/3})(1+ {2\alpha_s\over 3\pi}) \ ,$$
which includes the lowest-order contribution in $\alpha_s$ (the exchange
energy).
As discussed in the preceding section, in 
hadronic models that are characterized
by Eqs.~(\ref{eq:asymp1a}) and (\ref{eq:asymp1b}), the quadratic terms 
dominate at high densities, so that
\begin{eqnarray}
{\lim_{{\cal E}\to \infty}} 
 {\cal E}/ \rho \propto \rho \ ,
 \label{eq:neuteb1}
\end{eqnarray}
for $\zeta\geq 0$ and $\xi=0$.
This is also true if $\zeta=0$ and $\xi\geq 0$.
Thus, at sufficiently high densities, neutron matter always has higher
energy compared to the quark phase \cite{BAYM76}, and the two phases can be
connected by a Maxwell construction, which signals the transition from 
hadron to quark matter.

The situation is different in the general case $\zeta>0,\, \xi>0$.
For the asymptotic form of the energy corresponding to Eq.~(\ref{eq:asymp2}),
one obtains
\begin{eqnarray}
 {\lim_{{\cal E}\to \infty}} {\cal E}/ \rho =
 {3\over 4}(3\pi^2)^{1/3}\left[1+\Bigl({2\over \pi^2\zeta}\Bigr)^{1/3}
                +{1\over 2}\Bigl({1\over \pi^2\xi}\Bigr)^{1/3}\right]
		\rho^{1/3}+O(\rho^{-1/3}) \ ,
		\label{eq:neuteb2}
\end{eqnarray}
which is, up to the prefactor, the same leading behavior as in 
Eq.~(\ref{eq:bageb}) for the quark phase.
Therefore a phase transition is possible,
{\em i.e.}, neutron matter has higher energy, only if
\begin{eqnarray}
 3^{1/3}\left[1+\Bigl({2\over \pi^2\zeta}\Bigr)^{1/3}
                +{1\over 2}\Bigl({1\over \pi^2\xi}\Bigr)^{1/3}\right]
		> f(\alpha_s) \ .
		\label{eq:condition}
\end{eqnarray}
This remarkable observation implies
that for sufficiently large values of the nonlinear couplings,
the matter remains in the hadron (neutron) phase, at least in 
the simple model discussed here.
More generally, one observes that independent of the asymptotic form, 
increasing the couplings $\zeta$ and $\xi$ increases the density at which
the phase transition occurs (if it does), since the hadronic EOS becomes
softer as the nonlinear couplings increase.

The different possibilities are illustrated in Fig.~\ref{fig:f10}.
The curve labeled $a$ corresponds to Eq.~(\ref{eq:neuteb1});
the asymptotic behavior is clearly different from the quark EOS.
The phase transition occurs at roughly $3\rho_0$, and the Maxwell 
construction is indicated by the dotted line.
In the situation described by curve $b$, neutron matter and quark matter
have a similar asymptotic behavior, but the condition (\ref{eq:condition}) 
remains true, and the two curves cross, leading to a phase transition at
roughly $6\rho_0$.
Finally, curve $c$ lies completely below the quark EOS and a transition
is not possible; neutron matter is stable at all densities.

If one introduces interaction terms of higher than fourth order in the 
fields, for example, $(V_{\mu}V^{\mu})^3$, the hadronic energy is dominated 
by the Fermi gas contribution, and 
Eq.~(\ref{eq:neuteb2}) must be replaced by
\begin{eqnarray}
 {\lim_{{\cal E}\to \infty}} {\cal E}/ \rho =
 {3\over 4}(3\pi^2)^{1/3}\rho^{1/3}+O(\rho^{-1/3}) \ .
		\label{eq:neuteb3}
\end{eqnarray}
Since $3^{1/3} \!<\! f(\alpha_s)$, no transition is 
possible in the asymptotic regime for any choice of 
hadronic parameters in this case.
On the other hand, the transition regime also depends on the model and
parameters used for the description of the quark phase. 
In our model, the vacuum constant $b$ and strong coupling $\alpha_s$
can be used to shift the transition point substantially \cite{BDS87,BETHE87},
so that a transition may occur outside the asymptotic regime.
It is clear, however, that the strong model dependence in the hadronic EOS
introduces large uncertainties in any attempted prediction of these values.

The consequences for neutron stars in this model follow straightforwardly.
Nonlinear vector meson interactions soften the hadronic EOS, which lowers
the maximum neutron star mass and increases the density of the transition 
to quark matter.
In contrast, a stiff hadronic EOS lowers the density of the phase transition,
and since the quark matter EOS is soft, also tends to decrease the maximum
star mass.
Thus it may be impossible to decide, from neutron star masses alone, whether
quark matter cores exist in neutron stars, and similar conclusions may be
drawn about other exotic phenomena that soften the EOS.
(The situation is complicated further by the continuous nature of the
transition when two conserved charges are involved, which is the more
physical case \cite{GLENN92}.)
Whereas it might be possible, using the results of more advanced calculations
of the finite-density hadron/quark phase transition, to rule out certain
parametrizations of the hadronic EOS, existing uncertainties in both the
nature of the phase transition and in the high-density hadronic EOS preclude
any definite conclusions at this time.

%

%
\section{Summary}
In this paper we study the equation of state of nuclear and neutron-star
matter based on relativistic mean-field theory.
Our starting point is an effective action (or energy functional) containing 
Dirac nucleons and local scalar and vector fields.
These fields are interpreted as relativistic Kohn--Sham potentials, and
nonlinear interactions between the fields are introduced to parametrize
the density dependence of the energy functional.
We calibrate the energy functional by observing that at normal nuclear
densities, the ratios of the mean fields to the nucleon mass are small, and
thus the nonlinear interactions can be truncated at some low order in 
the fields.
The unknown parameters can then be fit to properties of nuclear matter near
equilibrium that are known to be characteristic of the observed bulk and
single-particle properties of nuclei.

We then extrapolate the resulting equation of state into the neutron-rich,
high-density regime to calculate the properties of neutron stars.
Two problems arise in the extrapolation:
First, even with a significant truncation of the energy functional, the
unknown parameters are underdetermined.
Thus there exist families of parameters that reproduce exactly the same
nuclear matter properties near equilibrium, but which produce potentially
different high-density equations of state.
Second, terms omitted from the functional because they are negligible
at normal density may become important at densities relevant for neutron
stars.
This is true even if we assume that the coupling parameters are ``natural'',
which means that they are all of roughly the same size when expressed
in appropriate dimensionless ratios.

Our basic goal is to determine, in light of these two problems, whether the
calibration at equilibrium nuclear matter density is sufficient to predict
a maximum neutron star mass within a reasonably small range.
This is relevant in view of recent calculations that hope to see evidence
for ``new'' physics in neutron stars (such as quark cores, strange matter,
or kaon condensates) based on the need for a softer high-density equation
of state than that provided by neutrons, protons, and electrons alone.
These calculations assume that the high-density behavior of these
more mundane components is well known, and in particular, that the
contributions of many-nucleon forces are 
negligible\cite{GLENN92,MARU94,KNORR95,PRAK95a,PRAK95b,FUJII96}.
These many-body, density-dependent forces are precisely the ones that are
difficult to calibrate using observed nuclear properties; the question
is whether one can build nuclear equations of state with different types
of many-body forces that all reproduce the observed properties near
equilibrium, but which yield significantly different results at high density.

By beginning with a meson self-interaction potential containing arbitrary
powers of scalar-isoscalar ($\sigma$), vector-isoscalar ($\omega$), 
and vector-isovector ($\rho$)
fields, we show that the meson nonlinearities can have a profound
effect on the high-density equation of state.
In models where the vector mesons enter the potential at most quadratically,
the equation of state is stiff and asymptotically approaches $p = \cal E$.
(The Walecka model is a special case.)
In models where the vector fields enter with high powers, these fields become
negligible at high density, and the asymptotic equation of state resembles
that of a free, relativistic gas: $p = {\cal E}/3$.
The intermediate case occurs when the vector fields enter quartically; the
asymptotic equation of state is still soft ($p = {\cal E}/3$), but the
approach to the asymptotic limit is determined by the coupling parameters.

We illustrate these results using specific models containing quartic
$\omega$ and $\rho$ meson couplings and also sixth- and eighth-order
$\omega$ couplings.
All models are calibrated to exactly the same nuclear properties at
equilibrium, for all choices of parameters.
We find that by far the most important coupling constant is that of the 
quartic $\omega$ term; even when restricted by the requirements of 
naturalness, variations in this parameter can produce variations of 
nearly one solar mass in the predicted maximum neutron star mass.
(This is true for both pure neutron matter and for beta-stable matter.)
This uncertainty is clearly relevant on the scale on which one hopes to
identify new, exotic effects.
Moreover, increasing the strength of the quartic coupling softens the
equation of state, which is precisely the effect sought from the exotica.
The impact of a quartic $\rho$ meson coupling is smaller, and its effects
are only appreciable in stars made of pure neutron matter; the maximum
masses of stars computed with beta-stable matter show little change
when this parameter is varied within the bounds imposed by naturalness.
Similarly, sixth-order $\omega$ interactions have only a modest effect on
the predicted maximum mass, and by the time the eighth-order terms are
included, the high-density equation of state is already so soft that these
terms are negligible.

We emphasize that the importance of these many-body effects is not limited
to the domain of relativistic mean-field theories; equations of state
based on nonrelativistic potentials \cite{FP81}
use interactions that are calibrated
to few-body systems and that are insensitive to possible six- or eight-body
forces that may be relevant at high density.
This is especially important because a mean-field calculation with 
four-component spinors and just {\em two-body\/} Lorentz scalar forces 
already implicitly contains an infinite string of many-body forces if it is 
recast in terms of two-component spinors.

To illustrate these difficulties more concretely, we also study the role
of the hadron/quark phase transition in a simple two-phase model with a
first-order transition.
We find that with couplings well within the bounds of naturalness, it is
possible to push the phase transition to arbitrarily high density, and
even to make it disappear altogether.
Although the absence of a phase transition is probably unrealistic and 
could serve to exclude some values of the hadronic parameters, there are
still too many uncertainties on both sides of the transition (as well as
in the nature of the transition itself!) to make any definitive statements.

We therefore conclude that existing methods for calibrating the nuclear
equation of state for extrapolation into the neutron-rich, high-density
regime appropriate for neutron stars cannot constrain the predicted maximum
star mass well enough to make reliable statements about the existence of
``new'' physics beyond the dynamics of neutrons, protons, and electrons.
We show that the uncertainties arising from an incomplete knowledge of the
hadronic many-body forces are much larger than those arising from an 
imperfect knowledge of the properties of nuclear matter near equilibrium, 
such as the nuclear matter compression modulus.
Even the old question of the role of the hadron/quark phase transition
is problematic, since the high-density hadronic equation of state can be 
made essentially as soft as desired by the addition of nonlinear 
interactions that are still consistent with equilibrium nuclear matter 
properties.

One positive conclusion is that the most important nonlinear parameter
is that of the quartic $\omega$ interaction.
If this term could be accurately calibrated, the uncertainties introduced
by other nonlinear interactions are likely to be tolerable.
(One caveat: we do not study carefully the sensitivity to variations in
mixed scalar--vector interactions because of the overwhelming sensitivity
to the quartic $\omega$ term; if the latter were well constrained, the
role of scalar--vector interactions should be examined in more detail.)
Although this coupling has not been extensively studied in mean-field
calculations, there are several possibilities for determining it reasonably
well.
First, since a quartic $\omega$ interaction leads to a nonlinear density
dependence in the vector part of the baryon self-energy, one could
calibrate this interaction by fitting to the self-energy obtained in a
Dirac--Brueckner--Hartree--Fock calculation, for example.
Although some initial attempts at this procedure have been made
\cite{GMUCA92}, the 
resulting parameters are not always natural; it is probably necessary to
repeat the procedure using all possible scalar and vector self-interactions
through fourth order and to fit both the scalar and vector part of the
self-energy simultaneously.
Second, the nonlinearities in the energy functional can be interpreted in 
terms of effective masses for the vector and scalar mesons (defined by
diagonalizing the matrix of appropriate second derivatives of the energy
functional).
This may provide useful constraints in the future, if concrete empirical
information on these effective masses becomes available.
Third, it is possible that additional observables in finite nuclei could
constrain the nonlinear interactions.
For example, some recent work suggests that the ratio of the
nuclear matter ``skewness''
(which is related to the third derivative of the energy with respect to
density at equilibrium) to the compression modulus $K$ is constrained by
monopole vibrations.
Although a recent calculation of nuclear ground-state properties shows
little correlation with this ratio \cite{FST96}, a more detailed examination
of dynamical effects could provide meaningful constraints.

To summarize, precise predictions of the properties of neutron stars
apparently require more accurate calibrations of the nuclear equation of
state than are currently available.
It is especially important to have the high-density behavior of the
``standard'' components (neutrons, protons, and electrons) under control
before one can make reliable statements about the existence of ``new''
physics.
Since the window on experimentally observable nuclear properties is a narrow
one, producing an equation of state that can be extrapolated with
confidence remains a major challenge.


\acknowledgments

We thank R. J. Furnstahl and H.-B.\ Tang for useful comments.
This work was supported in part by the Department of Energy
under Contract No.\ DE--FG02--87ER40365.

\newpage
%

%
%
\section*{Figure captions}
\global\firstfigfalse
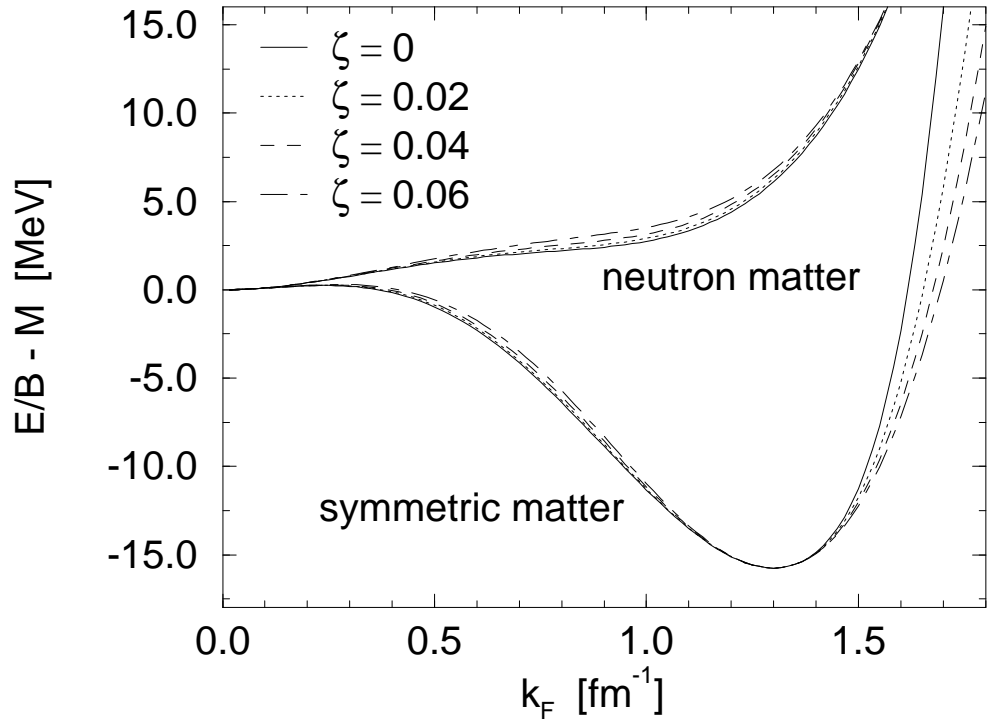
\begin{figure}[tbhp]
\caption{
Binding energy of symmetric and neutron matter.}
\label{fig:f1}
\end{figure}
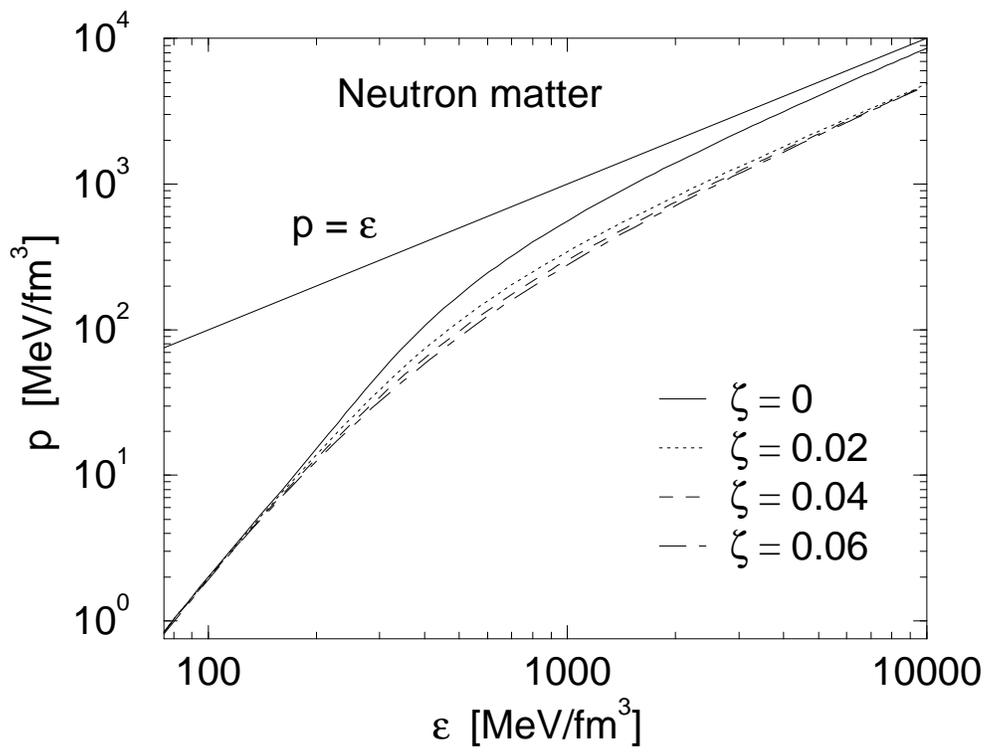
\begin{figure}[tbhp]
\caption{
Equation of state of neutron matter.}
\label{fig:f2}
\end{figure}
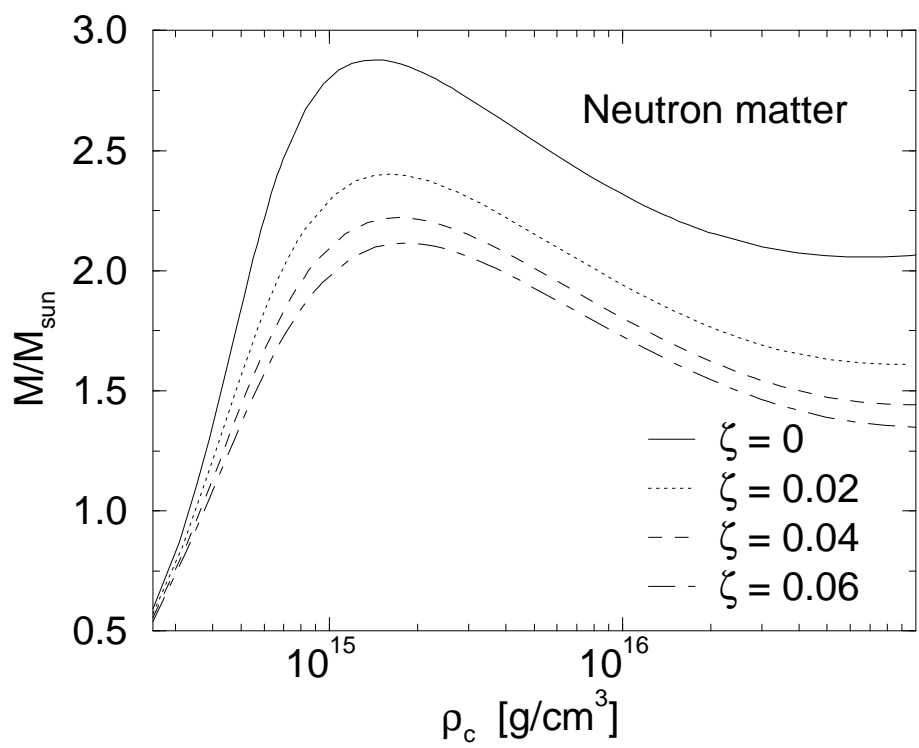
\begin{figure}[tbhp]
\caption{
Neutron star masses for different nonlinear couplings $\zeta$.
Pure neutron matter is assumed.}
\label{fig:f3}
\end{figure}
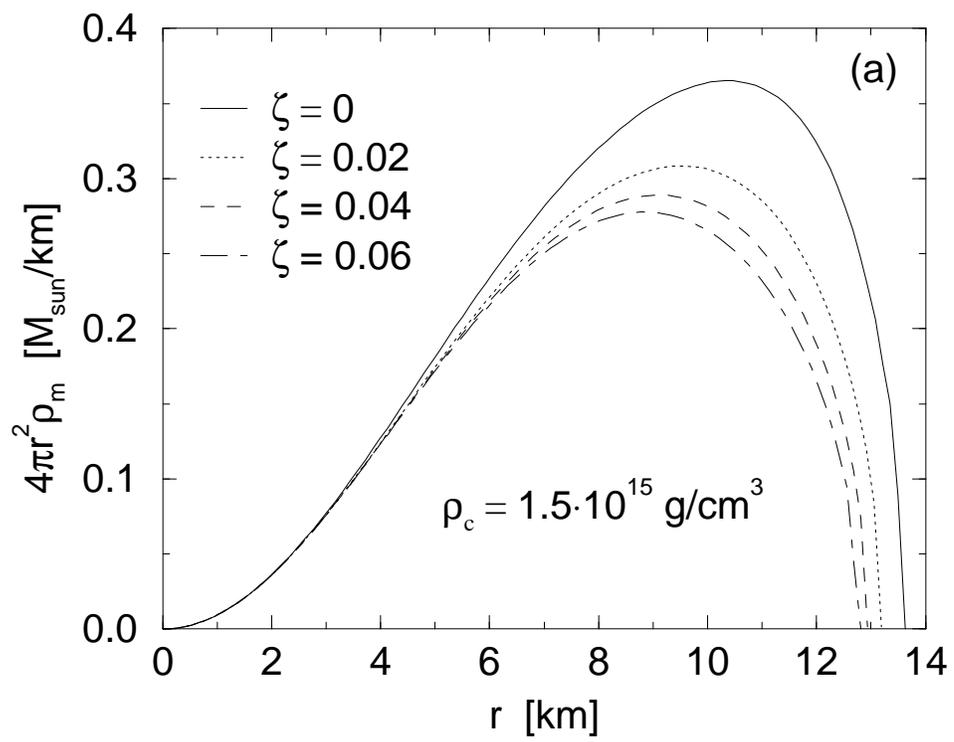
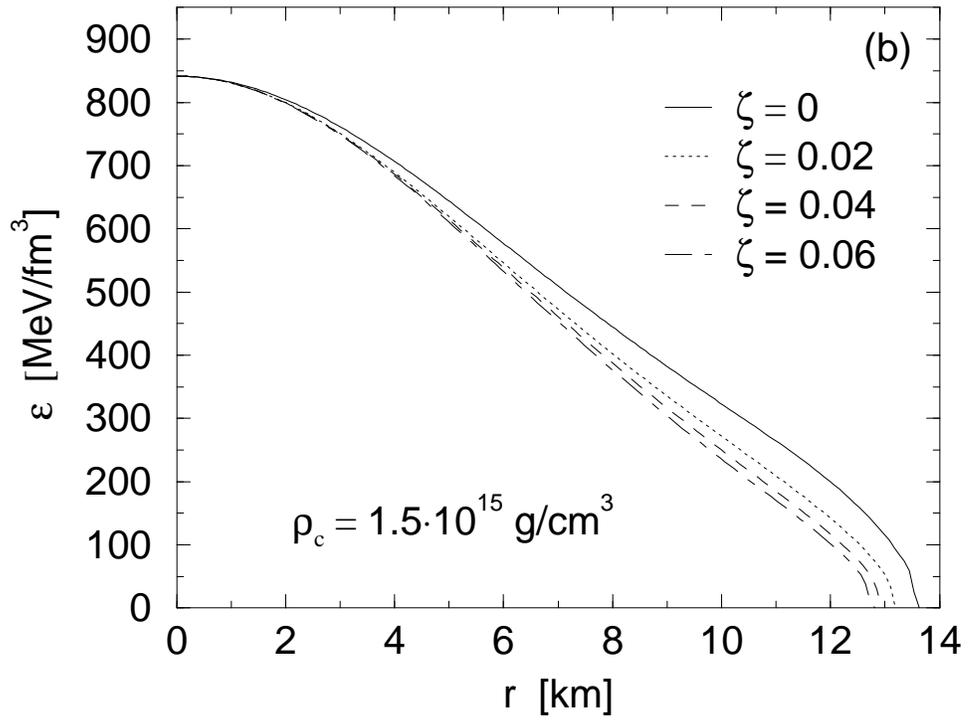
\begin{figure}[tbhp]
\caption{
Density distributions of neutron stars as a function of the radius,
calculated with different parameters $\zeta$. 
The central mass density $\rho_c = 1.5\times 10^{15}$ g/cm$^3$ is the same
for all curves.
Part (a) shows the radial mass densities and part (b) the corresponding 
energy densities. The stars acquire most of their mass from the region 
$100 \protect\lesssim {\cal E} \protect\lesssim 500$ MeV/fm$^3$.}
\label{fig:f4}
\end{figure}
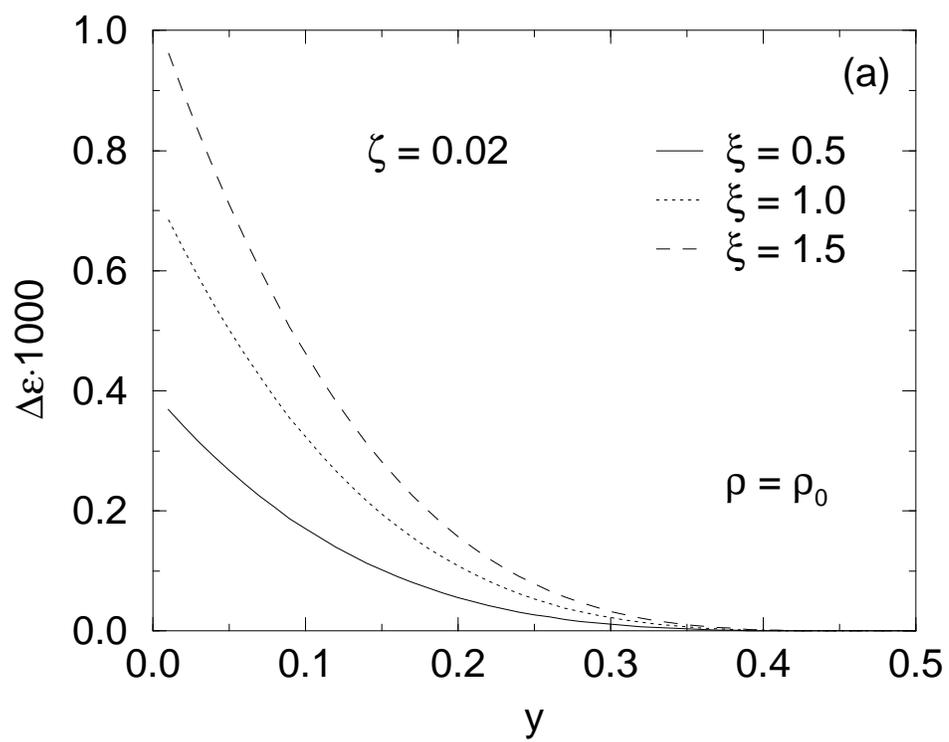
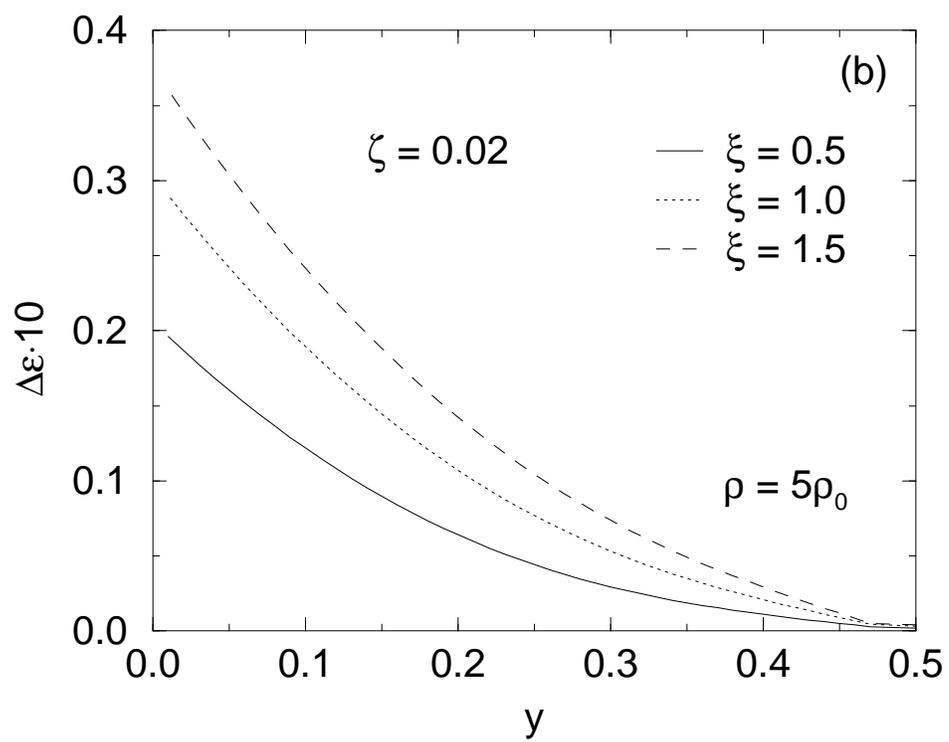
\begin{figure}[tbhp]
\caption{
Energy difference
$\Delta {\cal E} = [{\cal E}(\xi=0)-{\cal E}(\xi)]/{\cal E}(\xi=0)$
of nuclear matter as a function of the proton fraction $y$
calculated at constant baryon density. In part (a) the density is fixed
at its equilibrium value $\rho_0$ and part (b) shows the results at
$5\rho_0$.
Note the different vertical scales in (a) and (b).
}
\label{fig:f5}
\end{figure}
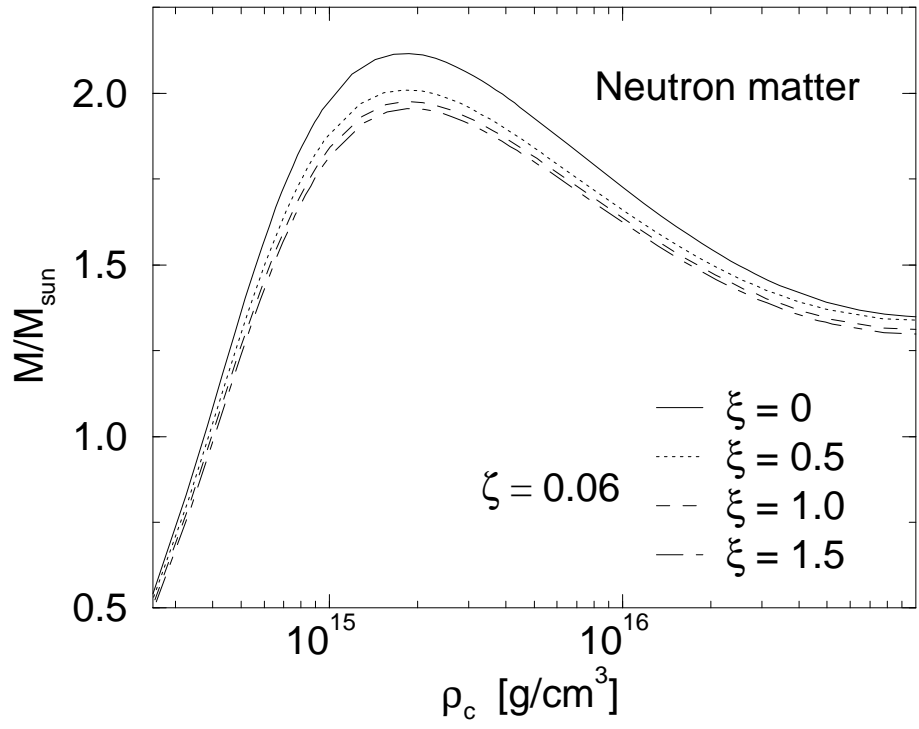
\begin{figure}[tbhp]
\caption{
Neutron star mass for different values of $\xi$ at fixed $\zeta$.
Pure neutron matter is assumed.}
\label{fig:f6}
\end{figure}
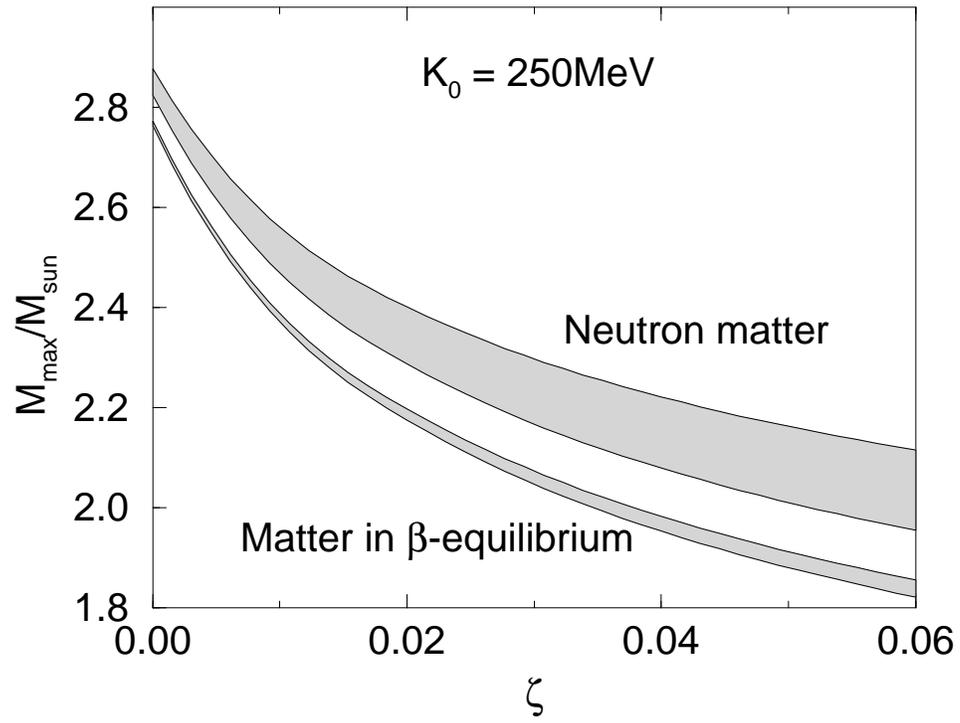
\begin{figure}[tbhp]
\caption{
Maximum neutron star mass as function of $\zeta$ and $\xi$.
Results for pure neutron matter 
and for matter in $\beta$-equilibrium are displayed.
The shaded areas show the mass range obtained when
$\xi$ is varied; the upper boundaries correspond to
$\xi = 0$ and the lower boundaries to $\xi = 1.5$.
}
\label{fig:f7}
\end{figure}
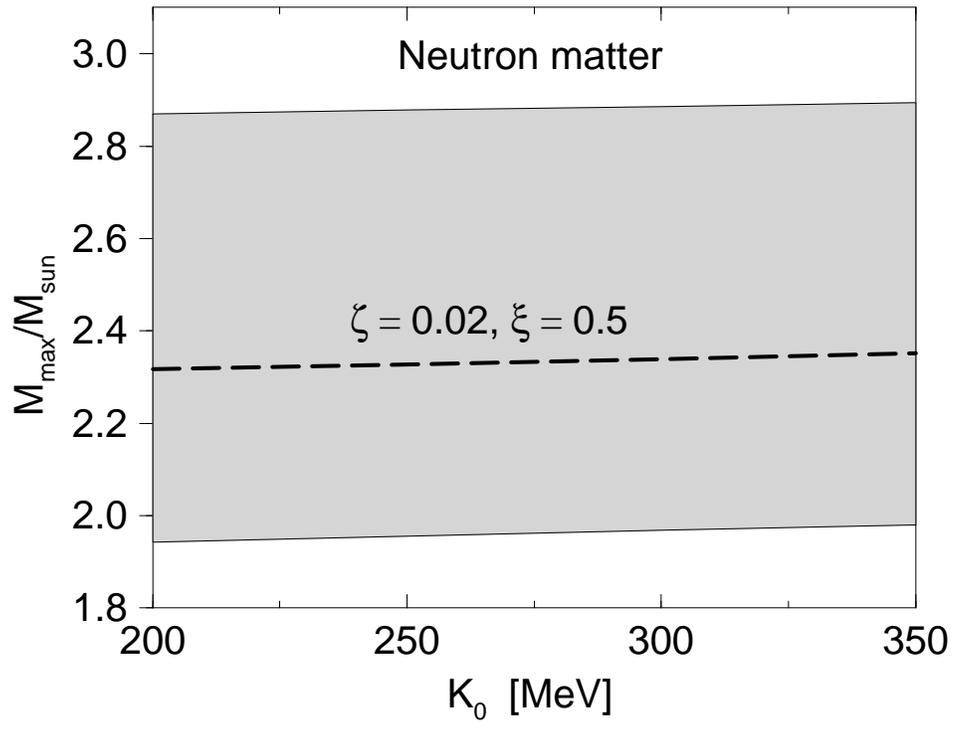
\begin{figure}[tbhp]
\caption{
Maximum neutron star mass as a function of the compression modulus $K_0$.
(All other nuclear matter inputs are held fixed.)
The shaded area marks the covered range of masses. The upper boundary
corresponds to $\zeta = 0,\, \xi=0$ and the lower boundary to
$\zeta = 0.06,\, \xi=1.5$. For fixed values of the nonlinear couplings,
the mass changes marginally with the compression modulus.
This can be seen
from the dashed line inside the shaded area, which corresponds to 
$\zeta=0.02,\, \xi=0.5$
}
\label{fig:f8}
\end{figure}
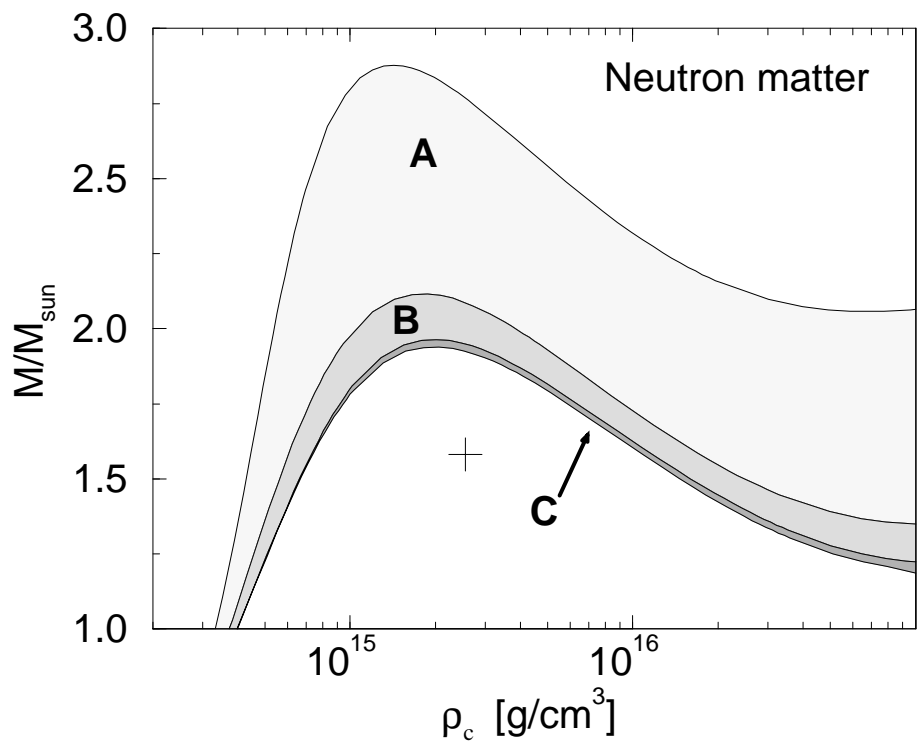
\begin{figure}[tbhp]
\caption{
Neutron star masses for models with different nonlinear
couplings for the neutral vector meson.
The uppermost curve corresponds to the Walecka model, including
cubic and quartic couplings for the scalar meson. 
Region {\em A} shows the range of masses obtained when the quartic vector 
coupling is turned on.
Regions {\em A} and {\em B} correspond to models with up to sixth-order 
terms, and regions {\em A}, {\em B}, and {\em C} include an eighth-order 
term.
The values of the nonlinear couplings are chosen within a natural range,
as described in the text.
The cross indicates the maximum mass ($1.58\, M_\odot$) obtained for
beta-stable matter in a calculation that includes the $\zeta$, $\zeta'$,
$\zeta''$, and $\xi$ couplings (see the text).
}
\label{fig:f9}
\end{figure}
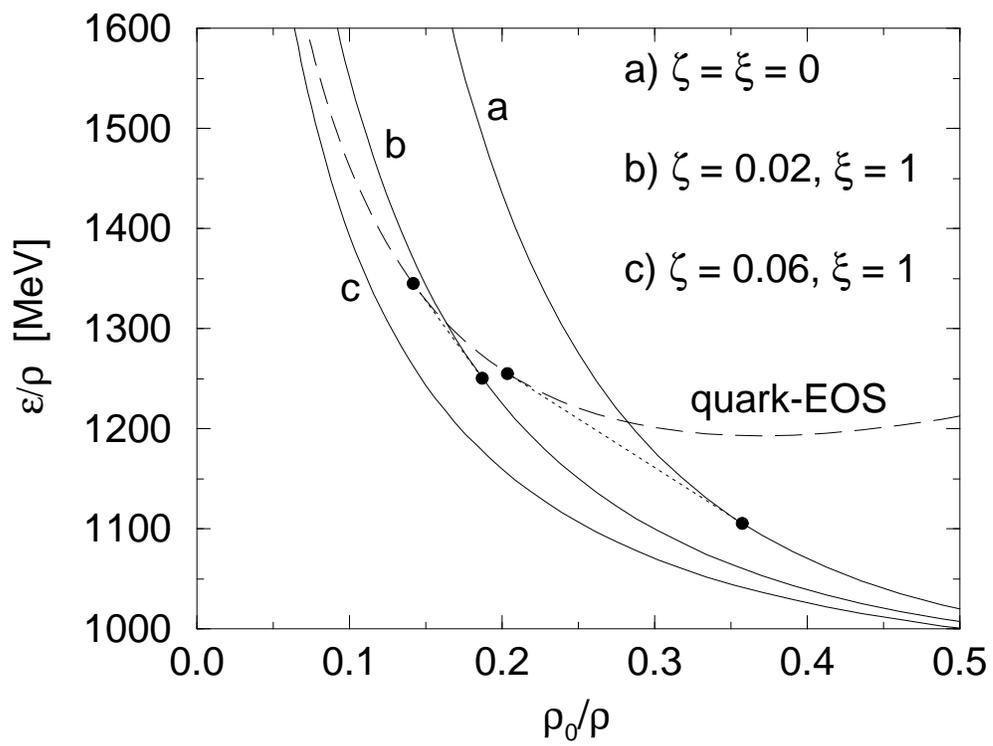
\begin{figure}[tbhp]
\caption{
Energy per baryon for neutron matter (solid) and quark matter
(dashed).
The quark matter results are calculated using $b = 120$ MeV/fm$^3$ and
$\alpha_s = 0.4$ \protect\cite{BDS87}.}
\label{fig:f10}
\end{figure}
\newpage
\begin{table}[tbhp]
\caption{Equilibrium Properties of Nuclear Matter}
\medskip
\begin{tabular}[b]{cccccc}
$(k_{\scriptscriptstyle\rm F})_0$ & $\rho_{\scriptscriptstyle 0}$ 
                & $M_0^* /M$ &
                $e_{\scriptscriptstyle 0}$ & $K_0$ & $a_4$ \\
\hline
1.30\,fm$^{-1}$ & 0.1484\,fm$^{-3}$ & 0.60 & $-15.75$\,MeV & 250\,MeV &
					 35\,MeV\\
\end{tabular}
\label{tab:one}
\end{table}

\end{document}